%% file: Main.tex
\documentclass[pra, twocolumn,amsmath, amssymb, notitlepage, longbibliography, showpacs, superscriptaddress]{revtex4-2}

\usepackage{graphicx}
\usepackage{dcolumn}
\usepackage{bm}
\usepackage[breaklinks=true]{hyperref}
\usepackage{upgreek}

\hypersetup{
	colorlinks   = true, 
	urlcolor     = blue, 
	linkcolor    = blue, 
	citecolor   =  red 
}

\newcommand{\ppcKet}[1]{\ensuremath{\left|{#1}\right\rangle}}
\usepackage{tikz}
\usetikzlibrary{calc, spy}
\usepackage[american]{circuitikz}
\usepackage{pgfplots}
\usetikzlibrary{external}
\begin{document}

\preprint{APS/123-QED}

\title{Qubit-controlled directional edge states in waveguide QED}%





\author{Prasanna Pakkiam}
\affiliation{ARC Centre of Excellence for Engineered Quantum Systems, School of Mathematics and Physics, The University of Queensland, Saint Lucia, Queensland 4072, Australia}

\author{N. Pradeep Kumar}
\affiliation{ARC Centre of Excellence for Engineered Quantum Systems, School of Mathematics and Physics, The University of Queensland, Saint Lucia, Queensland 4072, Australia}

\author{Mikhail Pletyukhov}
\affiliation{Institute for Theory of Statistical Physics, RWTH Aachen University, 52056 Aachen, Germany}

\author{Arkady Fedorov}
\affiliation{ARC Centre of Excellence for Engineered Quantum Systems, School of Mathematics and Physics, The University of Queensland, Saint Lucia, Queensland 4072, Australia}

\date{\today}

\begin{abstract}
We propose an \emph{in-situ} tunable chiral quantum system, composed of a quantum emitter coupled to a waveguide based on the Rice-Mele model (where we alternate both the on-site potentials and tunnel couplings between sites in the waveguide array). Specifically, we show that the chirality of photonic bound state, that emerges in the bandgap of the waveguide, depends only on the energy of the qubit; a parameter that is easy to tune in many artificial atoms. In contrast to previous proposals that have either shown imperfect chirality or fixed directionality, our waveguide QED scheme achieves both perfect chirality and the capability to switch the directionality on demand with just one tunable element in the device. We also show that our model is easy to implement in both state-of-the-art superconducting circuit and quantum dot architectures. The results show technological promise in creating long-range couplers between qubits while maintaining, in principle, zero crosstalk.

\end{abstract}

\maketitle



{\it Introduction --} Engineering novel interactions between distant quantum emitters mediated by photons travelling in a 1D waveguide is crucial for building large scale quantum networks \cite{Kimble2008}. In particular, realising chiral light-matter interaction that results in unidirectional emission and scattering of photons can enable routing of quantum information between different nodes \cite{Cirac1999} and can also aid in on-chip integration of non-reciprocal devices such as single-photon diodes, transistors, circulators and amplifiers \cite{Lodahl2017,RosarioHamann2018, sun2018single, Wanjura2020}. Furthermore, such chiral interactions has far reaching applications in probing novel complex many body quantum states \cite{Ramos2016, Pichler2015, Bello2019}. Remarkable experimental progress has also been made in realizing such interactions in a variety of waveguide quantum electrodynamics (wQED) platforms. More specifically, in nanophotonic waveguides such as nanofibers and photonic crystals, spin-momentum locking between quantum emitters and the guided modes has led to the observation of asymmetric spontaneous emission of photons \cite{Sllner2015, Mitsch2014}. 

Distinctly different chiral quantum phenomenon has also been demonstrated in superconducting qubits coupled to a 1D lattice that realizes the photonic analog of the SSH model \cite{Su1979,Bello2019, Kim2021}. Here, the presence of qubit acts as a domain wall and thereby breaks the chiral symmetry of the chain. When the qubit energy lies in the bandgap of the waveguide, it induces a photonic bound states that is akin to an edge state and decays to either ends of the waveguide depending on the location of the qubit in the unit-cell. Note that these edge states are static photonic wavefunctions as opposed to moving currents seen in other edge states such as those in 2D topological insulators. Several proposals has also been reported to achieve $in-situ$ tunable chiral photonic states which would then enable on-demand routing of quantum information in a network. To this end, a wQED platform based on a giant atom coupled to Josephson metamaterial has been proposed in \cite{Wang2021}. Here, the chiral bound state stems from the interference due to the non-local interaction induced by the giant atoms at different location in the waveguide. However, in order to flip the direction of the photon decay, one has to either tune the coupling between the qubit and the waveguide or dynamically alternate the impedance of the waveguide which is cumbersome in an experimental set-up.

In this work we propose a different chiral quantum system that utilizes just one frequency tunable qubit to switch the chirality of the (either photonic or electronic) bound state on-demand. In our model the waveguide is implemented by periodically modulating the on-site potentials and the hopping energy between the sites. Such a waveguide resembles the Rice-Mele model that can support both uni-directional and bi-directional edge states at different energies~\cite{PhysRevLett.49.1455}. Furthermore we show that the directionality of these edges states can be switched by simply tuning the transition frequency of an artificial atom coupled to a defect site in the chain. In contrast to the previous proposal of Ref.~ \cite{Wang2021}, our model offers the conditions where the states are perfectly directional with mathematically zero value of the wave-functions of the bound states on the `wrong' side of the waveguide. We provide a detailed analysis by considering realistic experimental conditions such as finite nature of the device as well coupling to measurement leads. Moreover, we suggest two possible experimental realization of out wQED model that can be readily implemented using superconducting quantum circuits and quantum dot devices.

In addition to directional bounds states, there has been active research to develop chiral photonic interactions using giant atoms~\cite{Guimond2020,arxiv.2212.11400}. The tunable nature of these interactions arises by utilizing the non-local system topology, the nonlinear nature of the emitters as well as parametrically modulating the coupling between the two emitters. These proposals are used to generate directional emission and absorption of photons in the passband of the waveguide which can potentially used for transfer of quantum states, as well as the generation and manipulation of stabilizer codes for quantum error correction. Our proposal use static bound states  in the stopband which has promise in realising long-range inter-qubit interactions while minimising crosstalk between adjacent qubits.

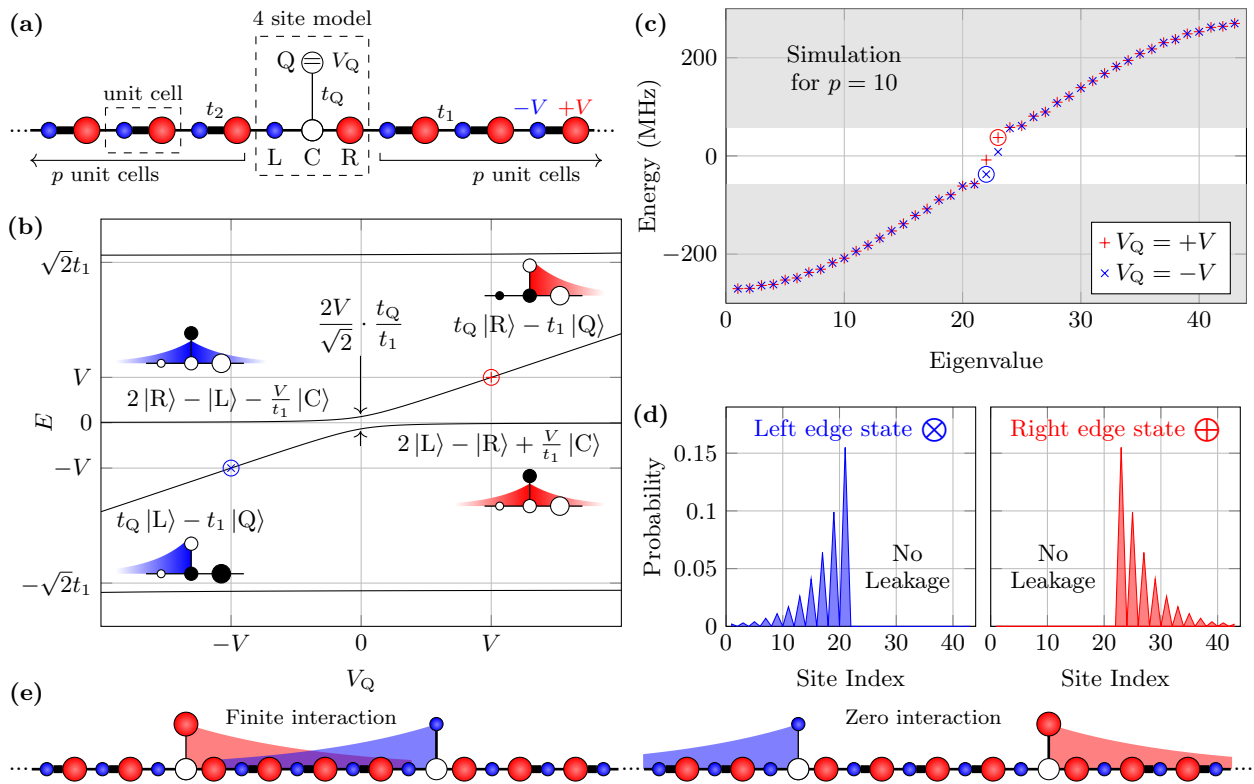
\begin{figure*}[!ht]
	\input{FigNumIdeal2}
	\caption{\label{fig:FigNumIdeal} \textbf{Proposed Rice-Mele Y-coupler for qubits}. The switching of the edge state is done by simply tuning the on-site potential of the qubit site. \textbf{(a)} Layout of the Rice-Mele coupler where a central site (in white) couples two Rice-Mele chains which have a modulating on-site potential (red and blue for $+V$ and $-V$) and modulating tunnel coupling ($t_1$ and $t_2$ with $t_2>t_1$). The qubit separately couples to the central site via tunnel coupling $t_\textnormal{Q}$ with itself having a tunable on-site potential $V_\textnormal{Q}$. We term each strongly coupled pair of sites a `unit cell' and the Rice-Mele chains each have $p$ unit cells. The sites for the effective 4-site model in Eqn.~\ref{eqn-Hamil-4-site-model} are labelled Q, C, L and R. \textbf{(b)} Energy spectrum of the proposed Rice-Mele Y-model, shown in (a) with $V=V_\textnormal{R}=-V_\textnormal{L}$. The eigenstates producing the leftward and rightward edge-states are $t_\textnormal{Q}\ppcKet{\textnormal{L}}-t_1\ppcKet{\textnormal{Q}}$ and $t_\textnormal{Q}\ppcKet{\textnormal{R}}-t_1\ppcKet{\textnormal{Q}}$ and occur at $V_\textnormal{Q}=-V$ and $V_\textnormal{Q}=V$ respectively (shown by the blue and red circles) as highlighted by sketched population on the two inset diagrams (with white for finite population while black for zero population). The remaining two eigenstates produce a bidirectional population spread as shown in the top-left and bottom-right inset diagrams. \textbf{(c)} Numeric simulation of the eigenvalues produced by the full coupler design described in Eqn.~\ref{eqn:fullHamilIdeal}. The simulation takes $V=37.5\,\textnormal{MHz}$, $t_1=120\,\textnormal{MHz}$, $t_2=150\,\textnormal{MHz}$ and $t_\textnormal{Q}=62.5\,\textnormal{MHz}$. There are $p=10$ unit cells on both Rice-Mele chains. The spectra are plotted for both $V_\textnormal{Q}=\pm V$. The non-shaded region indicates the band-gap which contains two gap states. The encircled edge states correspond to the operating points shown in (b). \textbf{(d)} Occupation probabilities exposing the edge-states when taking the respective eigenvectors from the spectrum in (c). Note that the occupation probability is zero on one side for both directional edge-states. The qubit site 44 has been omitted for clarity, while the site indices are enumerated from 1 to 43 from the most leftward site to the most righward site (so 22 is the central site). \textbf{(e)} An example multi-qubit array where multiple edge states can be made to either overlap (finite interaction) or have zero crosstalk by simply switching the qubits' on-site potentials.}
\end{figure*}

{\it Directional edge states in Rice-Mele waveguide--} Our goal is to create an in-situ tunable directional edge state that has all its population along one direction of the array while having zero population along the other direction. Before we discuss how to tune the directionality, it is useful to review the chiral edge states supported by the Su-Schrieffer-Heeger (SSH) model~\cite{PhysRevLett.42.1698}. As discussed in Appendix~\ref{appen-ssh}, an SSH chain is an array of sites with alternating tunnel couplings $t_1$ and $t_2$ (with $t_2>t_1$). This arrangement naturally introduces an energy spectrum that has two clusters of states with a large band gap in between. If every site has an adjacent site to which it is strongly tunnel coupled (that is, $t_2$), then no edge states occur and the band gap remains with no states. However, if there is a lonely site that is not tunnel coupled to another site with the stronger $t_2$, an edge state (rooting from the lonely site) with the desired directionality forms in the band gap. Many proposals to date have realised the edge states. However, the shortcomings are either in their sheer complexity (like giant atoms~\cite{Wang2021} or multiple qubits~\cite{Pichler2015}) or the inability to switch the direction of the edge state in-situ~\cite{Kim2021}.


Here, we propose a simpler approach that leads to the ability to switch the direction of the edge state by only varying the on-site potential of the tunable qubit site rooting the edge state. To create such a directional edge state, the intuitive approach is to merge two SSH chains to a central site. To point the edge-state along a given direction, we need to zero the population on the adjacent node on one side (to realise no edge-state in that direction like in Fig.~\ref{fig:figSSH}a) while letting the adjacent node on the other side have a non-zero population to spawn an edge-state (like in Fig.~\ref{fig:figSSH}b). However, the SSH model alone does not have enough degrees of freedom to tune the direction as there is no physical characteristic to differentiate both chains with respect to the central site. To circumvent this problem, we use a more general Rice-Mele model where the on-site potentials are alternated to break the inversion symmetry between both directions~\cite{PhysRevLett.49.1455}.

As shown in Appendix~\ref{appen-qubit-side-coupled-why}, the tunable qubit needs to be side-coupled to a central site which joins the two Rice-Mele chains as shown in Fig.~\ref{fig:FigNumIdeal}a. We couple the two Rice-Mele chains to the central site equally to have the resulting directional edge states symmetric when pointing leftward and rightward (see Appendix~\ref{sec:appen:elimRemCombs} for more details). Additionally, this coupling is chosen to be the weaker $t_1$ to maximise the population away from the centre (qubit). As before, the on-site potentials must be different to break the inversion symmetry of the structure to enable directional edge states.

To get insight into operation of this device it is sufficient to consider just the four central sites highlighted in Fig.~\ref{fig:FigNumIdeal}a via the Hamiltonian:

\begin{equation}\label{eqn-Hamil-4-site-model}
	\mathbf{H}_\textnormal{y-cell}=\begin{pmatrix}
	V_\textnormal{Q} & -t_\textnormal{Q} & 0 & 0 \\
	-t_\textnormal{Q} & V_\textnormal{C} & -t_1 & -t_1 \\
	0 & -t_1 & -V & 0 \\
	0 & -t_1 & 0 & V
	\end{pmatrix},
\end{equation}
where the basis states (for the labelled sites) are $\ppcKet{\textnormal{Q}}$, $\ppcKet{\textnormal{C}}$, $\ppcKet{\textnormal{L}}$ and $\ppcKet{\textnormal{R}}$. This approximation holds when the qubit state exists within a large band gap opened by the two Rice-Mele chains as discussed in Appendix~\ref{sec:appen:mainGreen}. Fig.~\ref{fig:FigNumIdeal}b summarises the resulting energy spectrum observed when sweeping $V_\textnormal{Q}$; the derivation of the features are given in Appendix~\ref{sec:appen:TcdomU}. We can switch between the leftward and rightward edge-states (sketched in the inset figures) $t_\textnormal{Q}\ppcKet{\textnormal{L}}-t_1\ppcKet{\textnormal{Q}}$ and $t_\textnormal{Q}\ppcKet{\textnormal{R}}-t_1\ppcKet{\textnormal{Q}}$ by rapidly sweeping $V_\textnormal{Q}$ past the anti-crossing at $V_\textnormal{Q}=0$. Note that the sweep rate must be faster than the anti-crossing gap: $\tfrac{2V}{\sqrt{2}}\cdot\tfrac{t_\textnormal{Q}}{t_1}$ with the gap closing when $t_\textnormal{Q}\to0$ where the qubit is completely decoupled. The other corresponding states on the anti-crossing are bidirectional and lie near $E=0$ as seen by their finite components in both $\ppcKet{\textnormal{L}}$ and $\ppcKet{\textnormal{R}}$ (again sketched in the inset). The two states near $E=\pm\sqrt{2}t_1$ are a part of the band states whereupon the remaining states of the Rice-Mele chains begin to appear as shown later in Fig.~\ref{fig:FigNumIdeal}c.

Given the basic operation of the Y-structure, consider the full Hamiltonian of the Y-structure coupled to both Rice-Mele chains as described in Fig.~\ref{fig:FigNumIdeal}a is:

\begin{align}\label{eqn:fullHamilIdeal}
    \mathbf{H}_\textnormal{y} &=
    -t_\textnormal{Q}|M\rangle\langle N|
    +\tfrac{V_\textnormal{Q}}{2}|N\rangle\langle N|\nonumber\\
    &-\tfrac{V}{2}|N_L\rangle\langle N_L|+\tfrac{V}{2}|N_R\rangle\langle N_R| - t_1\sum_{m=2p}^{N_R}|m\rangle\langle m+1|\nonumber\\
    &+\mathbf{H}_\textnormal{RM}^{1,p}+\mathbf{H}_\textnormal{RM}^{M+1,p} + h.c.
\end{align}
where the Hermitian conjugate applies to all listed terms. Here the length of the Rice-Mele chain is $2p$. That is, $p$ is the number of pairs of sites coupled adjacently via $t_2$. The index of the central site is $M=2N+2$, with the adjacent site to the left being $N_L=M-1$, the adjacent site on the right being $N_R=M+1$, while the qubit site is on $N=4p+4$ (and incidentally the dimension of this Hamiltonian). Note that given $p$ pairs, the dimension of the space spanned by $\mathbf{H}^{m,p}_\textnormal{RM}$ is $2p$. The first two lines of the equation represent the central four sites interlinked via the tunnel couplings $t_1$ and modulating on-site potentials $\pm V$ (the factor of a half is due to the Hermitian conjugate). The qubit is tunnel-coupled via $t_\textnormal{Q}$ to the central site and has an on-site potential of $V_\textnormal{Q}$. The Hamiltonians for the two interlinking Rice-Mele chains connecting to the central four sites are given via the tridiagonal matrix (when adding the Hermitian conjugate):

\begin{align}
   \mathbf{H}_\textnormal{RM}^{n,p}&=\frac{V}{2}\sum_{l=1}^{2p}(-1)^{l}|n+l-1\rangle\langle n+l-1|\nonumber\\
   &-t_2\sum_{l=1}^{p}|n+2l-2\rangle\langle n+2l-1|\nonumber\\
   &-t_1\sum_{l=1}^{p}|n+2l-1\rangle\langle n+2l|.
\end{align}
Here, the first line represents the modulating on-site potentials, while the last two lines represent the modulating tunnel couplings.

A numeric simulation of the Hamiltonian given in Eq.~(\ref{eqn:fullHamilIdeal}) is shown in Fig.~\ref{fig:FigNumIdeal}c with 10-cell Rice-Mele chains. The eigenspectrum shows the band-gap along with the two expected gap states. When switching between $V_\textnormal{Q}=\pm V$, the edge-state flips direction as seen by observing the corresponding eigenstates in Fig.~\ref{fig:FigNumIdeal}d. Note that the edge-states are unidirectional with \textit{exactly} zero probability on one side. The edge-state only has non-zero probability from the qubit-site onwards as expected from the 4-site model. This is in contrast to the tunable chiral quantum system proposed in \cite{Wang2021} where the chirality is not perfect and there is a finite probability of finding the photon on both the directions. Finally Fig.~\ref{fig:FigNumIdeal}e shows how we can exploit the switchable edge states to have adjacent either qubits interact when the edge states face each other like the two on the left~\cite{Kim2021}. Similarly, adjacent qubits can have zero crosstalk when their edge states face away from one another like the two on the right.



 {\it Influence of measurement ports --}The previous section introduced a Y-configuration that enables complete directional toggling of population along either side of the Rice-Mele chain. When running experiments to verify the presence of the edge states, the typical measurement will involve coupling the two edges of the chain to measurement ports. The influence of the ports on the resulting edge-states must be properly understood and thus, we apply the Green's function transport formalism~\cite{mesoDatta,PhysRevB.105.115419}. First one has to write down the Hamiltonian for the Y-configuration with some finite Rice-Mele chains and then to add the non-hermitian self-energy terms $\Sigma_n$ to the on-diagonal terms on sites $n$ to account for the influence of the measurement ports. To see this explicitly, consider $\mathbf{H}_\textnormal{y}$ with ports attached to the left most side $1$ and the right most site $N$:

\begin{equation}
    \mathbf{H}'_\textnormal{y}=\mathbf{H}_\textnormal{y}-i \Gamma (| 1 \rangle \langle 1 | + | N \rangle \langle N |),
\end{equation}
where we take the couplings $\Gamma$ to be equal on both ports. The corresponding Green's function as a function of the energy $\hbar\omega$ is:

\begin{equation}
    \mathbf{G}(\omega)=(\omega\mathbf{I}_{N}-\mathbf{H}')^{-1},
\end{equation}
where $\mathbf{I}_{N}$ is the identity matrix. Taking $G_{ab}$ to be the matrix component $(a,b)$ in $\mathbf{G}(\omega)$, we can write down the ports' transport transmission and reflection coefficients via the Fisher-Lee relations~\cite{fisherLee,mesoDatta}:

\begin{align}
    S_{ab}=\delta_{ab}-2i\Gamma G_{ab},
\end{align}
where we opt to use the reverse sign convention for clarity. Noting that the physical S-parameters (that is, measured ratio of RF signals or DC currents entering or leaving the ports~\cite{mesoDatta}) are given as $|S_{ab}|^2\textnormal{arg}(S_{ab})$, we can numerically calculate the physical scattering parameters via the above relations. Nonetheless, by exploiting symmetries, we provide exact analytic solutions for this Hamiltonian in Appendix~\ref{sec:appen:mainGreen}. As the solutions are algebraically cumbersome, we shall focus on the edge-states that occur at $\omega=V_\textnormal{Q}$ when $V_\textnormal{Q}=\pm V$. In this case, the transmission $S_{N1}=0$. The reflectances when the edge-state faces port 1 are:

\begin{align}
    S_{11}&=-1, &V_\textnormal{Q}&=-V\\
    S_{NN}&=-\frac{1+i\frac{2V}{\Gamma w_f}}{1-i\frac{2V}{\Gamma w_f}}, &V_\textnormal{Q}&=-V,
\end{align}
while the reflectances with the edge-state facing port $N$ are:

\begin{align}
    S_{11}&=-\frac{1+i\frac{2V}{\Gamma w_f}}{1-i\frac{2V}{\Gamma w_f}}, &V_\textnormal{Q}&=V\\
    S_{NN}&=-1, &V_\textnormal{Q}&=V
\end{align}
That is, the reflectance undergoes a relative $\pi$ phase shift and reflects completely when probing an edge-state directed towards the measured port. Note that in the isolated regime where $\Gamma\to0$, the reflectance of the opposite port is measured to remain at $1$, which is the expected value given that there is no population near this port. However, we see that as $\Gamma\to\infty$, the opposite port also yields the same phase in the reflected signal. That is, stronger coupling to the ports yields less chirality in the edge states as any macroscopic wavefunction will undergo greater losses to the ports. Note that $w_f$ is the measure of edge state wavefunction's decay from the beginning to end of the Rice-Mele chain. In the limit where the states are away from the edges of the band-gap, with the band-gap being large, we can approximate this term as:

\begin{equation}
    w_f\approx\frac{t_2^2-t_1^2}{t_2^2(\tfrac{t_2}{t_1})^{2p}-t_1^2}.
\end{equation}
Thus, for a small Rice-Mele chain where $p$ is small, $w_f$ tends to unity and the reflectance is limited by $\Gamma$. When $p$ is large, $w_f$ tends to zero as there is appreciable decay of the edge state leading to near zero population on the edge sites ($1$ or $N$). Thus, the $\Gamma$ term is overcome, yielding a lower port coupling, to give a reflection of $1$.

\begin{figure}[!ht]
	\input{FigGammaLDOS}
	\caption{\label{fig:FigGammaLDOS} \textbf{Numeric simulation demonstrating the edge-states produced by the Y-structure when probing the reflectance on port $1$ and $N$}. The simulation takes $V=37.5\,\textnormal{MHz}$, $t_1=120\,\textnormal{MHz}$, $t_2=150\,\textnormal{MHz}$ and $t_\textnormal{Q}=62.5\,\textnormal{MHz}$. There are $p=10$ cells of Rice-Mele chains. The leftward edge-state is initialised via $V_\textnormal{Q}=-V$. \textbf{(a)} Here $\Gamma$ is varied and the left panel shows the trajectory of $S_{NN}$ on the Argand plane on increasing $\Gamma$. For four selected values of $\Gamma$, the resulting LDOS are plotted alongside (the qubit site is omitted in this plot for clarity). It is clear that the chirality of the wavefunction is slowly eroded with increasing $\Gamma$ as population. In addition, the contrast used to distinguish the edge-state by comparing $S_{11}$ and $S_{NN}$ is also diminished. \textbf{(b)} Plot of associated chirality $\chi$ as the port coupling $\Gamma$ is increased. The resulting edge-state spawning near port $N$ causes a loss in chirality $\chi$.}
\end{figure}
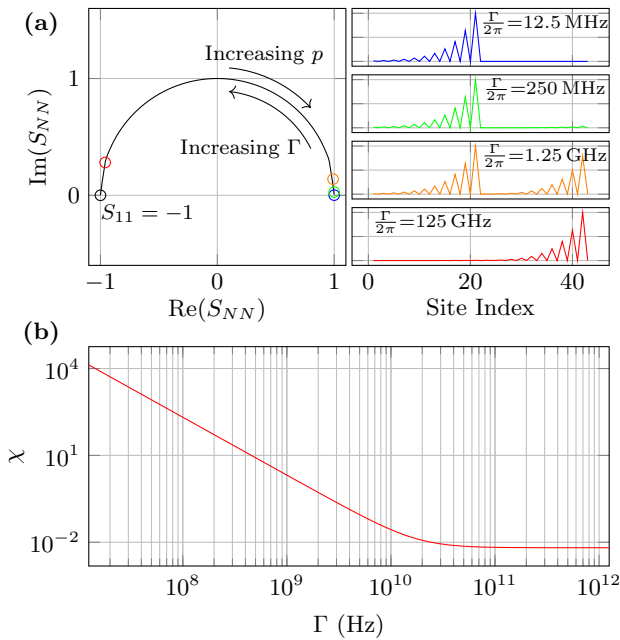

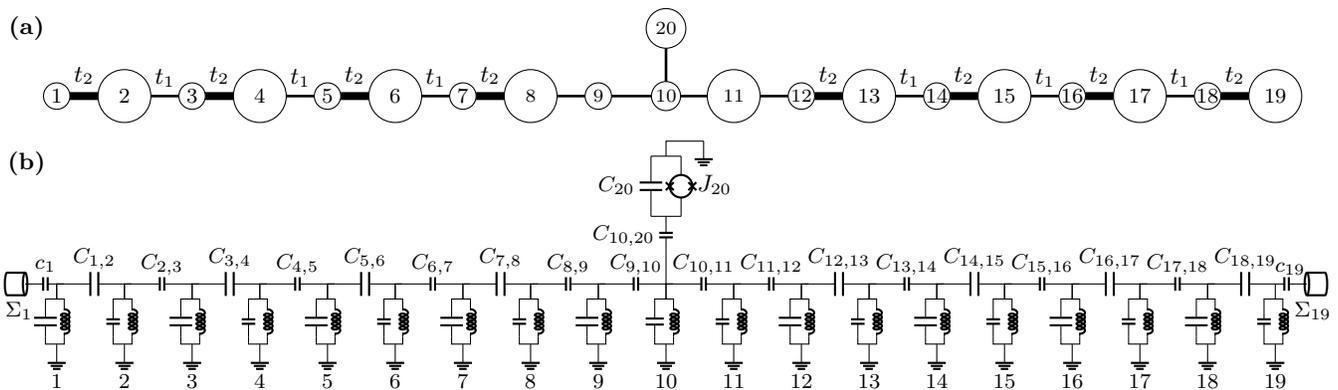
\begin{figure*}[!htb]
	\input{figCqedTranslation}
	\caption{\label{fig:figCqedTranslation} \textbf{Implementing the Rice-Mele Y-coupler using cQED elements}. The individual sites are formed with resonators while the tunnel couplings are implemented with coupling capacitors. \textbf{(a)} A simple coupled using chains with $p=4$ cells. Site 20 has a tunable on-site energy. \textbf{(b)} Implementing the structure shown in (a) using cQED elements. Each site $m$ has a resonator comprising of a capacitor $C_m$ and inductor $L_m$ (not labelled for clarity). The resonant frequencies alternate with the on-site potentials (shown via the larger and smaller capacitances for smaller and larger on-site potentials respectively). The tunnel couplings are implemented via capacitors $C_{m,n}$. The capacitances $C_{m,n}$ are also shown to alternate to account for the modulating $t_1$ and $t_2$ (being smaller and larger capacitances respectively). The tunable element is implemented via a flux-tunable Transmon qubit. If this structure is not coupled to other qubits and is just being measured via two ports, the circuit will contain two transmission lines $\Sigma_1$ and $\Sigma_{19}$ with coupling capacitors $c_1$ and $c_{19}$.}
\end{figure*}

Fig.~\ref{fig:FigGammaLDOS}a shows the changing $S_{NN}$ when probing an edge-state facing port $1$. It is once again clear that when $\Gamma$ is increased, the reflectance moves about the semi-circle from $S_{NN}=1$ to $S_{NN}=-1$. In doing so, the local density of states (LDOS) shifts (details of its calculation are given in Appendix~\ref{appen:LDOS}) from a strongly chiral edge-state with all population exclusively on the left hand side to being on sides. The interpretation is that when a strongly coupled port makes site $N$ starts to dominate the adjacent tunnel coupling $\Gamma\gg t_2$ (thereby, making a large energy cost to occupy site $N$), the site $N-1$ starts to become isolated from site $N$. Thus, like in Fig.~\ref{fig:figSSH}, an edge-state forms from site $N$. A similar edge state does not form from site $2$ as it is effectively zeroed on this site by the main edge-state from the central site. We can now define a simple chirality factor:

\begin{equation}
	\chi=\frac{\sum_{n=1}^{M-1}P_n}{\sum_{n=M+1}^{N}P_n},
\end{equation}
where $P_n$ is the occupation probability at site $n$ given the LDOS. By definition, $\chi=\infty$ when there are no ports. When increasing $\Gamma$ to move from $S_{NN}=1$ to $S_{NN}=-1$, the resulting change in chirality is shown in Fig.~\ref{fig:FigGammaLDOS}b. The chirality drops quadratically until it settles at a steady-state value whereupon the central edge-state and the edge-state at port $N$ settle to the limit where the port coupling fully isolates site $N-1$.


{\it Implementation -- }The proposed structure for a switchable edge state can be implemented either as photonic edge-states in circuit quantum electrodynamics (cQED) or electronic edge-states in conventional quantum dot system. This section highlights the details in design and implementation for both platforms.

{\it Implementing in cQED--}
The Y-structure and the associated Rice-Mele chains can be implemented using cQED elements. An example implementation of a $p=4$ Y-structure in Fig.~\ref{fig:figCqedTranslation}a is shown in Fig.~\ref{fig:figCqedTranslation}b. We implement each site $n$ with an $LC$-oscillator (consisting of $C_n$ and $L_n$) where its resonant frequency corresponds to the on-site potential. The tunnel-couplings are achieved via capacitors connecting across the resonators ($C_{n,n'}$ across resonators or sites $n$ and $n'$). The tunable side-coupled site (in this example, site 20) is achieved via a flux tunable Transmon qubit. The resulting edge states are photonic edge states across multiple resonators. As discussed in Appendix~\ref{sec:appen:cQED}, the corresponding on-site potentials $V_n\in\{V,-V\}$ and tunnel-couplings $t_{n,n'}\in\{t_1,t_2,t_\textnormal{Q}\}$ are:

\begin{align}
    V_n&=\frac{\hbar}{\sqrt{L_nC_{n(B)}}}-\hbar\omega_0\\
    t_{n,n'}&=-\frac{\hbar\sqrt{Z_nZ_{n'}}}{2C_{n(B)}C_{n'(B)}}C_{n,n'},
\end{align}
where $C_{n(B)}$ is the sum total of all capacitances connected to site $n$ and $Z_n=\sqrt{C_{n(B)}/L_n}$ with the important assumption: $C_{n,n'}\ll C_{n(B)}$. Since the Rice-Mele chain requires a modulation of positive and negative energies, we globally offset the on-site energies by $\hbar\omega_0$. Note that $\hbar\omega_0$ vertically recenters the spectrum in Fig.~\ref{fig:FigNumIdeal}b. Choosing a global nominal inductance $L_n=L_0$, we can solve for $C_{n(B)}$. Thus, we obtain the $C_{n,n'}$ for a given $t_{n,n'}$. Note that $C_{n,n'}$ is linearly proportional to $t_{n,n'}$ as shown in Fig.~\ref{fig:figCqedTranslation}. On obtaining all the $C_{n,n'}$ terms, we can finally obtain the resonator capacitances $C_n$.

Given that the cQED implementation will likely be tested via probing ports on the end sites (sites 1 and 19 in the example given in Fig.~\ref{fig:figCqedTranslation}), it is useful to understand the parameters concerning the coupling of the array to these ports. As discussed in Appendix \ref{sec:appen:cQED}, the port couplings can be modelled via the self-energy term given by:

\begin{equation}
    \Sigma_n=\Delta\omega_n+i\frac{\kappa_n}{2},
\end{equation}
where the frequency shift and photon loss rate are given as:

\begin{align}
    \Delta\omega_n&\approx\frac{\omega_n}{\sqrt{1+\frac{c_n}{C_n}}}-\omega_n\\
    \kappa_n&\approx\frac{c_n^2R}{C_n^2L_n},
\end{align}
Note that the idea is that the lead connecting to site $n$ has a resistance $R$ (typically $50\,\Omega$) and a coupling capacitor $c_n$ satisfying the limit $c_n\ll C_n$. Additionally, note that $\Gamma_n=\kappa_n/2$. Appendix~\ref{appen-cQED-SPICE} shows the SPICE simulations done on the model shown in Fig.~\ref{fig:figCqedTranslation} to verify the expressions for the individual lumped elements.

Due to the fixed values in the inductors and capacitors, it is important to consider the effect of fabrication defects on the final chiralities. As discussed in Appendix~\ref{appen-cQED-SPICE}, numerical simulations using a modest fabrication precision yielding $1\%$ variation in $V$, $t_1$ and $t_2$, yields a spread of $\chi=1130\pm80$ for a typical port-coupled implementation. In the case of no port-couplings (that is, coupling the chains to adjacent qubits like in Fig.~\ref{fig:FigNumIdeal}e), the median chirality drops from infinity to be bounded by the 5th and 95th quantiles as: 48000 ($6\times10^3$,$6\times10^6$). The sustained chirality confirms the fact that the states in the band-gap of a Rice-Mele or SSH waveguide are protected from localised defects as the wavefunction is macroscopically spread over multiple sites to effectively average out the effect of defects.

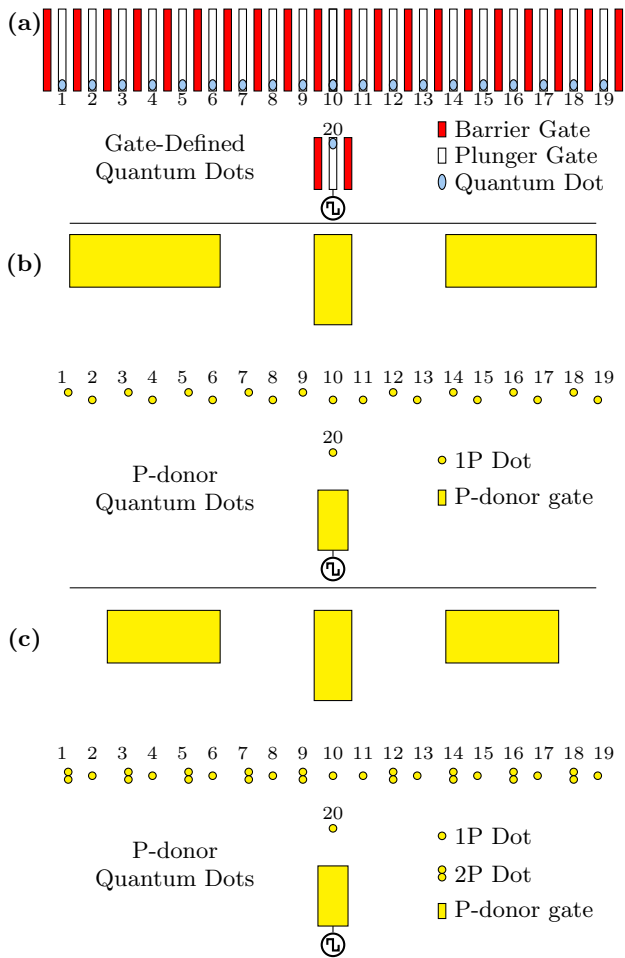
\begin{figure}[!ht]
	\input{figQuantumDots}
	\caption{\label{fig:qDots} \textbf{Implementing the Rice-Mele Y-coupler using quantum dots}. In each case, the outlined gates have constant voltage biases that are tuned once on startup. To rapidly control the direction of the bound state on demand, we highlight the single gate that requires DC pulsing.  \textbf{(a)} Implementation using gate-defined quantum dots. The $p=4$ structure is given in Fig.~\ref{fig:figCqedTranslation}a. The confinement potential formed by the gates traps the electrons into small quantum dots. The barrier gates can be used to in-situ tune the confinement potentials, while the plunger gates enable in-situ tuning of the on-site potentials. Thus, one may tune the parameters to form the required Rice-Mele chains. \textbf{(b)} When using STM patterned Si-P quantum dots, there is no need for dedicated confinement gates. The tunnel couplings are set by varying the distances between the P-donor dots (dots closer together had larger tunnel couplings), while the on-site potentials are tuned in-situ via the P-donor gates on the top (dots closer to the gates are tuned more strongly). Only the gate closest to site 20 is reservoir-coupled to the dot and is used to load electrons into the array. \textbf{(c)} The STM-patterned structure can be modified to provide alternating on-site potentials via larger 2P-donor dots. In this case, the on-site potentials dominate the tunnel couplings, whereupon we utilise the large $T_1$ and $T_2^*$ times of electron spins in P-donor dots. This regime of operation is discussed in Appendix~\ref{sec:appen:UdomTc}.}
\end{figure}

{\it Implementation in quantum dots--}
Given that the structure given in Eqn.~\ref{eqn:fullHamilIdeal} describes a network of tunnel-coupled sites with individual on-site potentials, a natural implementation falls directly in quantum dots. The chiral bound states are now electronic wavefunctions as opposed to the photonic wavefunctions seen when implementing in cQED. The required array is compatible with all major quantum dot platforms.

In the case of gate-defined quantum dots in Fig.~\ref{fig:qDots}a (such as SiGe or CMOS), the tunable tunnel couplings and on-site potentials enable a fully configurable array that can account for local defects~\cite{Hensgens2017,Mills2019}. Note that the outlined gates require constant DC biases, provided by DC looms, that can be rapidly tuned once on startup~\cite{Mills2019,Moon2020,Zwolak2020,Ziegler2022,RevModPhys.95.011006}. Afterwards, the direction of the bound state is rapidly controlled via DC pulses, sent through wide-band coaxial lines, on the central gate controlling to the qubit.

Another approach is to use atomic precision STM (scanning tunnelling microscope) patterned Si-P quantum dots where single P-donors are placed within a silicon substrate~\cite{Fuechsle2012,Hill2015,Kiczynski2022}. The P-donor dots yield a trapping potential without the need for confinement gates. To alternate the tunnel-couplings the distances between the dots are alternated with shorter distances used for the higher tunnel coupling $t_2$. As shown in Fig.~\ref{fig:qDots}b, the on-site potentials can be alternated by staggering the alternate dots closer to tuning gates. As dots are typically spaced in the order of $12\,\textnormal{nm}$ to create tunnel-couplings in the order of $10\,\textnormal{GHz}$, the on-site potentials only need to be tuned to approximately $1\,\textnormal{GHz}$ or approximately $4\,\upmu\textnormal{eV}$. Thus, the gates need to be only tuned approximately $100\,\upmu\textnormal{V}$ for typical lever-arm $\alpha$ values.

An alternative approach using P-donor quantum dots utilises the ability to tune the depth of the on-site potentials by placing another P-donor in the dot (a 2P cluster) like in Fig.~\ref{fig:qDots}c. In this case, depending on the positions of the P-donors in the Si crystal, the on-site potentials can vary in the order of $1\,\textnormal{meV}$ or $200\,\textnormal{GHz}$~\cite{Weber2014}. In such a case, the tunnel-couplings can be feasibly set in the order of $50\,\textnormal{GHz}$~\cite{Pakkiam2018}. Thus, the on-site potentials dominate the tunnel couplings as in the case discussed in Appendix~\ref{sec:appen:UdomTc}. In such a case, the edge-state can be tuned to be leftward, rightward or completely localised to the central qubit dot. Although this configuration was discouraged earlier, in the case of Si-P dots, the large electron spin $T_1$ and $T_2^*$ times may make this a desirable configuration~\cite{Pla2013}. That is, the electron spin can be localised and made to interact with neighbouring qubits on demand with zero crosstalk to other adjacent qubits.

It should be noted that the outlined proposals for the quantum dot structures require extra tuning gates (albeit, minimal with the Si-P implementations) compared to typical multi-qubit architectures. However, these are simply DC tuning gates that are only required to be tuned once to achieve the long-distance coupling with zero crosstalk between adjacent qubits. Whereas, a proposal that uses a normal quantum dot array through which to shuttle electrons (to mitigate crosstalk via distance) will require multiple fast-pulse gates~\cite{Fujita2017}. One notes that fast-pulse gates are more spatially expensive for they are extra coaxial lines in the dilution fridge as opposed to a compact DC wire loom.

{\it Conclusion -- }We have shown a general structure that can realise directional edge states with perfect chirality and in-situ switching via Rice-Mele chains. The direction can be switched by simply tuning the on-site potential of a single site; a feat that is easily realised in both cQED (via flux tuning) and quantum dot implementations (via gate voltage tuning). We show how our model can be implemented in cQED via a universal translation recipe that can be used to implement arbitrary site models using cQED elements. In addition, we show that the Rice-Mele chains can be implemented in quantum dot arrays in both the gate-defined and atomically defined quantum dot platforms. Finally, we provide a complete analysis of the influence of measurement probes. That is, although our model has zero crosstalk when coupling adjacent qubits, we show that there is a marginal loss in chirality when coupling the array to measurement probes in the case of verifying the directionality of a single qubit edge state. The overall simplicity of design and implementation shows promise in realising long-range inter-qubit interactions while minimising crosstalk between adjacent qubits.

\begin{acknowledgments}
The authors were supported by the Australian Research Council Centre of Excellence for Engineered Quantum Systems (EQUS, CE170100009). We also acknowledge Martin Maurer for insightful discussions.
\end{acknowledgments}

\section*{Competing Interests}

The Authors declare no Competing Financial or Non-Financial Interests.

\section*{Data Availability}

The datasets generated during and/or analysed during the current study are available from the corresponding author on reasonable request.

\section*{Author contributions}

The four-site model was developed by P. Pakkiam and M. Pletyukhov. The numeric simulations were handled by P. Pakkiam and N. Pradeep Kumar. Development of the Green's function formalism to handle the influence of ports was done by M. Pletyukhov. The formalism to implement in cQED was developed by P. Pakkiam, N. Pradeep Kumar and A. Fedorov. The project was supervised by A. Fedorov.

\appendix

\section{Edge states in SSH models}\label{appen-ssh}

\begin{figure}[!ht]
	\input{figSSH}
	\caption{\label{fig:figSSH} \textbf{SSH model giving rise to an edge-state}. The plots show the eigenspectrum while the inset figures show the corresponding site model. The thin lines indicate an inter-site tunnel coupling $t_1$, while the thicker lines indicate a stronger inter-site tunnel coupling of $t_2$. \textbf{(a)} An SSH model with $N=16$ unit-cells (32 sites in total). The model gives rise to a energy band gap. \textbf{(b)} A state arises in the gap (shown in blue) when adding a site on the end with tunnel coupling $t_1$ (this site does not belong to a unit-cell where it is tunnel-coupled $t_2$ to an adjacent site). The sketched plot above the inset figure shows the corresponding relative probability distribution of the resulting edge-state eigenstate. \textbf{(c)} Multiple edge states can be spawned via side-coupled sites (as shown on sites 10 and 15 which create). The two edge states, with population sketched in blue and green respectively, form two states in the band gap. The direction of the resulting edge state is not tunable for it only depends on whether the site to which it couples if an odd or even site.}
\end{figure}
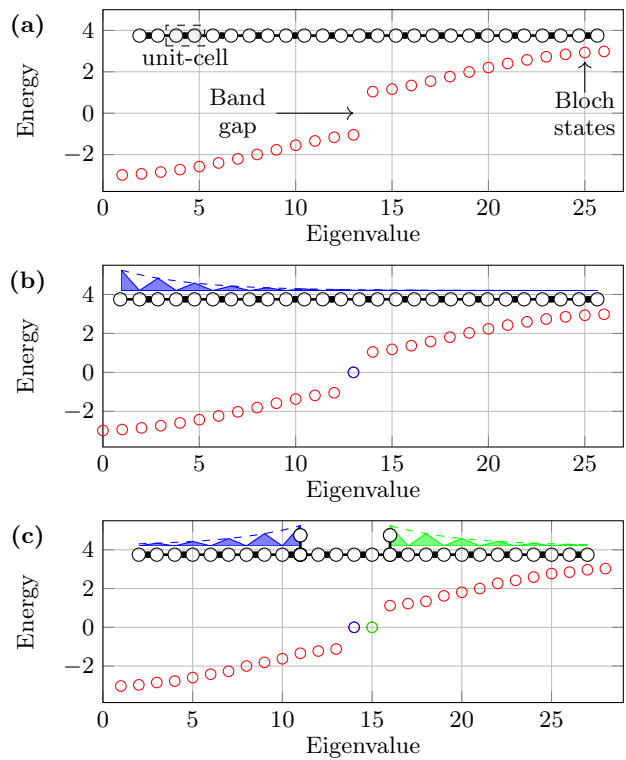

We consider the following Hamiltonian that represents the SSH mode1:

\begin{equation}
    \mathbf{H}_\textnormal{SSH} = -\sum_{n=1}^N t_1|2n\rangle\langle 2n+1|+t_2|2n\rangle\langle 2n-1|+h.c. 
\end{equation}
It consists of $2N$ lattice sites with alternating inter-site tunnel couplings $t_1$ and $t_2$ with $t_2>t_1$. The eigenspectrum of such a hamiltonian has an energy bandgap as shown in Fig.~\ref{fig:figSSH} for a chain consisting of $2N=16$ sites. A localized photonic state can be induced whenever we add a `lonely' site that is not paired up with an adjacent site by the stronger tunnel-coupling $t_2$ on either ends of the lattice:
\begin{align}
    \mathbf{H}^\textnormal{edge}_\textnormal{SSH} &= -t_1|0\rangle\langle1|\nonumber\\
    &-\sum_{n=1}^N t_1|2n\rangle\langle 2n+1|+t_2|2n\rangle\langle 2n-1|+h.c.
\end{align}
As shown in Fig.~\ref{fig:figSSH}b, such a configuration yields a state to appear in the middle of the energy band gap. This state is termed as edge state as it is localized at the unpaired edge site and decays into the bulk of the lattice. It has already been shown that an edge state can be created from any given site along an array of sites. In this case, the edge state has all its population on the given rooting site and sites only to one direction from that site. The edge state has zero population only along the opposite direction. Previous experiments have achieved this by introducing a qubit that is side-coupled to given lattice site. The resulting edge state roots from the side-coupled qubit and decays only in one direction with zero population on the other side as shown by the two side-coupled qubits yielding two states in the band-gap in Fig.~\ref{fig:figSSH}c. However, these edge states are fixed in direction and depend upon whether the qubit is coupled to an even or odd site~\cite{Kim2021}. Note that the direction is given by matching the side-coupled chain to Fig.~\ref{fig:figSSH}b. That is, the direction is given by the site that has a tunnel coupling of $t_2$ to continue the alternating $t_1$ and $t_2$ pattern.

\section{Why the qubit needs to be side-coupled}\label{appen-qubit-side-coupled-why}

\begin{figure}[!ht]
	\input{fig_straight}
	\caption{\label{fig:straight} Structure formed by merging two Rice-Mele chains at a tunable central site with on-site potential $V_\textnormal{C}$. The Rice-Mele chains consist of alternating on-site potentials $V_\textnormal{L}$ and $V_\textnormal{R}$ and alternating tunnel-couplings $t_1$ and $t_2$. The central site links to both chains via tunnel couplings $t_\textnormal{L}$ and $t_\textnormal{R}$. Note that $t_1\neq t_2$, $t_\textnormal{L}\neq t_\textnormal{R}$ and $V\neq 0$.}
\end{figure}
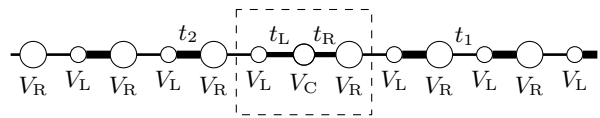

Consider Fig.~\ref{fig:straight} where a central site C is a qubit with a tunable on-site potential (for example, a flux-tunable transmon) is coupled to two Rice-Mele chains. To analyse this chain, we focus on the sites either side of the central site: L and R. The general Hamiltonian (taking tunnel-couplings on the left and right $t_\textnormal{L}$ and $t_\textnormal{R}$ and on-site potentials on the left, centre and right sites being $V_\textnormal{L}$, $V_\textnormal{C}$ and $V_\textnormal{R}$) is:

\begin{equation}
    \mathbf{H}_\textnormal{line}=\begin{pmatrix}
		V_\textnormal{L} & -t_\textnormal{L} & 0 \\
		-t_\textnormal{L} & V_\textnormal{C} & -t_\textnormal{R} \\
		0 & -t_\textnormal{R} & V_\textnormal{R}
    \end{pmatrix}.
\end{equation}
Now if this Hamiltonian were to form a directional edge-state, we must demand zero population on the sites L or R. That is, the eigenvectors of $ \mathbf{H}_\textnormal{line}$ are of the form $(a,b,0)$ and $(0,c,d)$. One trivial possibility to achieve that is to have zero couplings $t_\textnormal{L}$ or $t_\textnormal{R}$ on either side  of the zero site. As tunnel couplings are difficult to tune, we choose to discard this possibility. Another option is that the on-site potentials $V_\textnormal{L}$ and $V_\textnormal{R}$ are much greater than the tunnel couplings $t_\textnormal{L}$ or $t_\textnormal{R}$ to decouple the central cite from either left or right sites by tuning its energy in resonance with an opposite site (that is $\pm V$). As shown in Appendix~\ref{sec:appen:UdomTc}, this regime is not principally new as the toggling of the edge states is done so by completely transferring the coherent information solely onto the central site similar to protocols involving SWAP operations. 

\section{Case where on-site potential dominates tunnel-coupling}
\label{sec:appen:UdomTc}

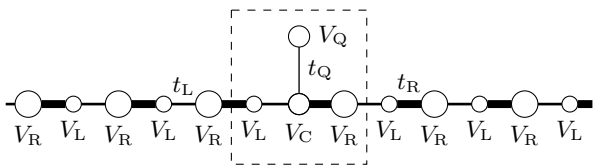
\begin{figure}[!ht]
	\input{fig_UdomTc}
	\caption{\label{fig:UdomTc} A general Rice-Mele Y-model with alternating on-site potentials $V_\textnormal{L}$ and $V_\textnormal{R}$ and alternating tunnel-couplings $t_\textnormal{L}$ and $t_\textnormal{R}$. The chains link to a central site with on-site potential $V_\textnormal{C}$. The central site links to a qubit (set at $V_Q$) with tunnel-coupling $t_\textnormal{Q}$. Note that $t_\textnormal{L}\neq t_\textnormal{R}$ and $V_\textnormal{L}=-V_\textnormal{R}$.}
\end{figure}

Now consider two general Rice-Mele chains shown in Fig.~\ref{fig:UdomTc}. In this section we investigate the case where the on-site potentials dominate the tunnel-couplings.

\begin{figure}[!ht]
	\input{fig_plotRM_UdomTc}
	\caption{\label{fig:plotRM_UdomTc} Plot of the 3 gap-state energy eigenvalues for the system shown in Fig.~\ref{fig:UdomTc} as a function of the qubit's on-site potential $V_\textnormal{Q}$. Note that $\ppcKet{\textnormal{C}}$ is off the plot as it was set to a large value to push it away from the band gap (the resulting perturbations slightly push the $\ppcKet{R}$ and $\ppcKet{L}$ downwards). Note that without loss of generality the plot takes $V_\textnormal{R}>V_\textnormal{L}$. The inset figures illustrate the edge-states that form (with white and black denoting large and small site populations respectively) along the $\ppcKet{\textnormal{L}}$, $\ppcKet{\textnormal{Q}}$ and $\ppcKet{\textnormal{R}}$ branches. Note that the $\ppcKet{\textnormal{Q}}$ has a strongly localised bound state.}
\end{figure}
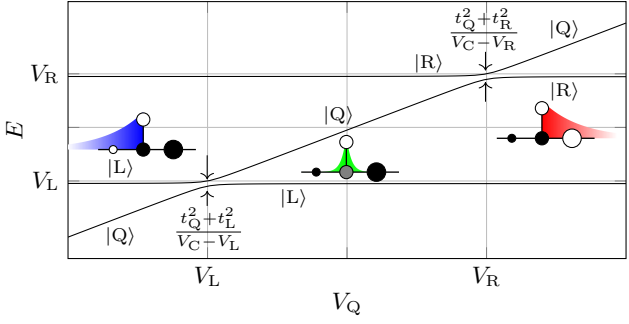

Assume that $t_\textnormal{L}$ and $t_\textnormal{R}$ are perturbative with respect to the rest of the Hamiltonian (this is fine as physically the two chains are weakly coupled to the central node). Now to suppress population in the central node C (as any directional edge-state should be symmetric in either the left or right configurations, the population here is irrelevant), one biases the energy of the central node such that any probability on that node results in an energy outside the gap. Away from anti-crossings at $V_\textnormal{Q}=V_\textnormal{L}$ and $V_\textnormal{Q}=V_\textnormal{R}$ (given by $t_\textnormal{L}$ and $t_\textnormal{R}$ respectively acting as the Pauli-$x$ term), the asymptotic eigenstates are simply given by the computational basis. To investigate each anti-crossing separately we take the resulting 3-level system (across the basis Q, C and L/R) with the L/R dot set to $\pm V_\textnormal{L}$ (taking $t_c$ to be $t_\textnormal{L}$ or $t_\textnormal{R}$ if investigating the L/R dots):

\begin{equation}
	\mathbf{H}=\begin{pmatrix}
		V_\textnormal{Q} & -t_\textnormal{Q} & 0 \\
		-t_\textnormal{Q} & V_\textnormal{C} & -t_c \\
		0 & -t_c & \pm V_\textnormal{S}
	\end{pmatrix}
	=\begin{pmatrix}
		V_\textnormal{S} & -t_\textnormal{Q} & 0 \\
		-t_\textnormal{Q} & V_\textnormal{C} & -t_c \\
		0 & -t_c & V_\textnormal{S}
	\end{pmatrix}.
\end{equation}
with the second equality centring the qubit energy at the anti-crossing and $\textnormal{S}\in\{\textnormal{R},\textnormal{L}\}$. The characteristic polynomial only involves the solving of a quadratic and one gets the lowest two eigenvalues and eigenvectors (corresponding to the anti-crossing) to be:

\begin{equation}
	\begin{cases}
		V_\textnormal{S} & \ppcKet{\textnormal{S}}-\frac{t_c}{t_\textnormal{Q}}\ppcKet{\textnormal{Q}}\\
		\frac{V_\textnormal{S}+V_\textnormal{C}-V_t}{2} & \ppcKet{\textnormal{S}}+\frac{t_c}{t_\textnormal{Q}}\ppcKet{\textnormal{Q}} - \frac{V_\textnormal{S}-V_\textnormal{C}+V_t}{2t_c}\ppcKet{\textnormal{C}}
	\end{cases},
\end{equation}
where $V_t=\sqrt{(V_\textnormal{C}-V_\textnormal{S})^2+4t_\textnormal{Q}^2+4t_c^2}$ and the associated eigenvectors are listed unnormalised for clarity. Note that the eigenvectors are simply the anti-symmetric and symmetric superpositions of the qubit and adjacent site states. Now the energy splitting at the anti-crossing is simply the subtraction of the two eigenvalues:

\begin{equation}
	\Delta_\textnormal{QS}\approx
	\frac{V_\textnormal{S}-V_\textnormal{C}+V_t}{2}
	\approx\frac{t_c^2+t_\textnormal{Q}^2}{V_\textnormal{C}-V_\textnormal{S}},
\end{equation}
where the last equality is simply a Binomial approximation given $V_C\gg t_c$. The resulting energy diagram for the three gap states is shown in Fig.~\ref{fig:plotRM_UdomTc}. Evidently, when $V_\textnormal{Q}\ll V_\textnormal{L}$, the eigenvalue approaches $\ppcKet{\textnormal{L}}$. To get to the purely rightward edge-state $\ppcKet{\textnormal{R}}$ in the region $V_\textnormal{Q}\ll V_\textnormal{L}$, one may adiabatically sweep $V_\textnormal{Q}$ through both anti-crossings. This operation effectively transfers the edge state purely into the qubit state. Note that in doing so, the resulting bound state on $\ppcKet{\textnormal{Q}}$ is very localised as the populations on two sites adjacent to the central site are close to zero.

\section{Case where tunnel-coupling dominates on-site potential}\label{sec:appen:TcdomU}

This section derives the energy diagram given in Fig.~\ref{fig:FigNumIdeal}b. The corresponding Hamiltonian of the central four sites, for the structure in Fig.~\ref{fig:FigNumIdeal}a, as discussed in the main text, is given as:

\begin{equation}
	\mathbf{H}_\textnormal{y-cell}=\begin{pmatrix}
	V_\textnormal{Q} & -t_\textnormal{Q} & 0 & 0 \\
	-t_\textnormal{Q} & V_\textnormal{C} & -t_1 & -t_1 \\
	0 & -t_1 & -V & 0 \\
	0 & -t_1 & 0 & V
	\end{pmatrix},
\end{equation}
where we take $V_\textnormal{R}=-V_\textnormal{L}=V$ (without loss of generality as any offset can be subtracted as an identity matrix term) and $V_\textnormal{C}=0$. In addition, $t_\textnormal{Q}$ is the tunnel-coupling of the qubit to the central site with $t_\textnormal{Q}\ll t_1$. In addition, we note once again that $t_1\gg V$ to realise that the on-diagonal terms act as a perturbation. The on-site potential terms cause an anti-crossing between the two inner energy eigenvalues. To find this energy-gap, take $V_\textnormal{Q}=0$ and observe the characteristic polynomial:

\begin{equation}
	t_\textnormal{Q}^2V^2-(2t_1^2+t_\textnormal{Q}^2+ 
	V^2)\lambda^2 + \lambda^4
	\approx	t_\textnormal{Q}^2V^2-2t_1^2\lambda^2+\lambda^4=0.
\end{equation}
The solutions are (for all combinations of plus and minus) $\lambda\pm\sqrt{t_1^2\pm\sqrt{t_1^4-t_\textnormal{Q}^2V^2}}$. Two of the solutions on taking very large $t_1$ yields the energies $\sim\pm\sqrt{2}t_1$. Similarly, on taking the Binomial approximation, the two remaining inner energies are given as:

\begin{equation}
	E_0\approx\pm\frac{t_\textnormal{Q}V}{\sqrt{2}t_1}.
\end{equation}
As discussed in the Appendix~\ref{sec:appen:elimRemCombs}, the edge-states are exactly purely directional when taking $V_\textnormal{Q}=\pm V$. The relevant characteristic polynomial in these positions is given as:

\begin{equation}
	(V\mp\lambda)(t_\textnormal{Q}^2V\pm(2t_1^2+t_\textnormal{Q}^2)\lambda\pm V^2\lambda\mp\lambda^3)=0.
\end{equation}
Clearly, $\lambda=\pm V$ are solutions; the associated purely directional eigenvectors are found to be  (unnormalised for clarity): $t_\textnormal{Q}\ppcKet{\textnormal{R}}-t_1\ppcKet{\textnormal{Q}}$ and $t_\textnormal{Q}\ppcKet{\textnormal{L}}-t_1\ppcKet{\textnormal{Q}}$ respectively. 

The remaining two branches can be found by noting that the characteristic polynomial, on dividing by $t_1$ and taking the appropriate limits $t_1\gg t_\textnormal{Q}$ and $t_1\gg V$, yields a solution $\lambda=0$ with the associated eigenvectors (unnormalised for clarity): $\ppcKet{\textnormal{L}}-\ppcKet{\textnormal{R}}+\tfrac{V}{t_c}\ppcKet{\textnormal{C}}$ and $\ppcKet{\textnormal{R}}-\ppcKet{\textnormal{L}}-\tfrac{V}{t_c}\ppcKet{\textnormal{C}}$, respectively. Note that the non-zero vector on taking the Hamiltonian matrix times either of these two eigenvectors is $\pm\tfrac{t_\textnormal{Q}V}{t_1}\ppcKet{\textnormal{Q}}$ and tends to zero as $t_c$ dominates both terms in the numerator.

Finally, we note that when we dynamically switch between states via rapid following. The transfer probability is given by the usual Landau-Zener-Stueckelberg-Majorana probability:

\begin{equation}
    P_\textnormal{LZSM}=\exp\left(-\frac{\pi}{h}\frac{E_0^2}{v}\right),
\end{equation}
where $v$ is the sweep rate (given as energy change per unit time $dV_\textnormal{Q}/dt$). To ensure $P_\textnormal{LZSM}\to1$, we take $v\gg E_0$, where we note that the leakage probability is exponentially suppressed with the sweep rate.

\section{Eliminating remaining combinations of tunnel-couplings and on-site potentials}\label{sec:appen:elimRemCombs}

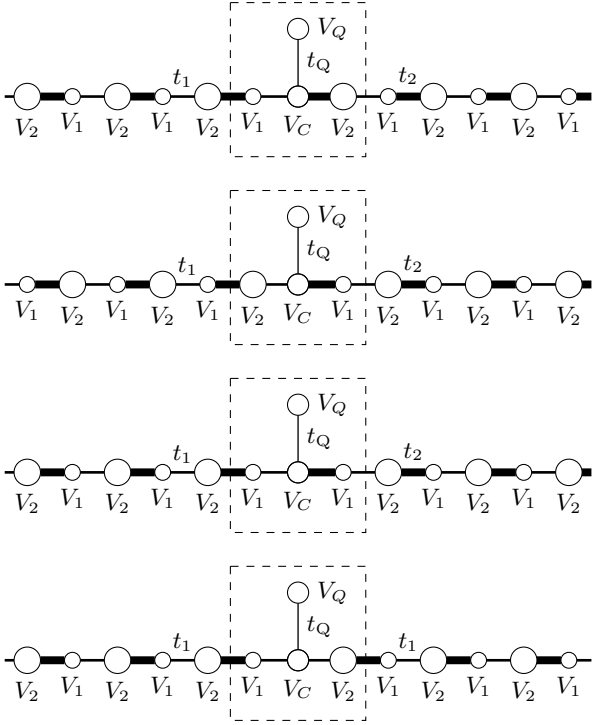
\begin{figure}[!ht]
	\input{fig_comb4}
	\caption{\label{fig:comb4} Four possible unique configurations in connecting the central qubit to two Rice-Mele chains. Note that in each configuration, we vary the positions of the on-site potentials $V_1$ and $V_2$ as well as the tunnel-couplings $t_1$ and $t_2$ connecting to the central site.}
\end{figure}

The discontinuity at the central site of the Rice-Mele chain shall be investigated as a four-site model. There is a choice in taking the values for the tunnel-couplings (selecting from $t_1$ and $t_2$) and on-site potentials (selecting from $V_1$ and $V_2$) on either side. There are clearly 16 combinations of which the 4 symmetric combinations are discarded for they will clearly not produce switchable edge-states as there is no difference in directing the photon exclusively on either branch. When discarding equivalent configurations in the remaining combinations, one arrives at those shown in Fig.~\ref{fig:comb4}.

Now before investigating all the cases, consider the most general Hamiltonian across the four sites in the centre across the basis Q, C, L and R:

\begin{equation}
	\mathbf{H}=\begin{pmatrix}
		V_\textnormal{Q} & -t_\textnormal{Q} & 0 & 0 \\
		-t_\textnormal{Q} & V_\textnormal{C} & -t_\textnormal{L} & -t_\textnormal{R} \\
		0 & -t_\textnormal{L} & -V & 0 \\		
		0 & -t_\textnormal{R} & 0 & V		
	\end{pmatrix},
\end{equation}
where we take $V_\textnormal{R}=-V_\textnormal{L}=V$ without loss of generality for any offset can be subtracted as an identity matrix term. Now we demand that there is some setting for $V_\textnormal{Q}$ such that the eigenstate is unidirectional. For a leftward vector, this is in general $\psi_\textnormal{L}=(a,b,c,0)$. This gives $\mathbf{H}\psi_\textnormal{L}=(bt_\textnormal{Q}+aV_\textnormal{Q},ct_\textnormal{L}+at_\textnormal{Q}+bV_\textnormal{C},bt_\textnormal{L}-cV,bt_\textnormal{R})$. To be an eigenvector this implies that $b=0$, $a=-c\cdot\tfrac{t_\textnormal{L}}{t_\textnormal{Q}}$ and $V_\textnormal{Q}=-V$. Thus, the associated leftward eigenstate is $t_\textnormal{Q}\ppcKet{\textnormal{L}}+t_\textnormal{L}\ppcKet{\textnormal{Q}}$. In a similar analysis, it can be shown that to have a completely rightward eigenstate, $V_\textnormal{Q}=V$ with the eigenstate $t_\textnormal{Q}\ppcKet{\textnormal{L}}+t_\textnormal{R}\ppcKet{\textnormal{Q}}$. It is immediately clear that in order to have a symmetric shape in the leftward and rightward eigenstates that one must set $t_\textnormal{R}=t_\textnormal{L}$ to thus, eliminate the first three possibilities shown in Fig.~\ref{fig:comb4}. This leaves the configuration discussed in the main text.

\section{Implementing site models in cQED}\label{sec:appen:cQED}

Consider an arbitrary network of sites that are tunnel-coupled to each other. We start by mapping each site $n$ to a capacitor $C_n$ and $L_n$ connected to ground. The electrical network of sites is interconnected from site $m$ to site $n$ via a capacitor $C_{m,n}$. We assume that $C_{m,n}\gg C_{p}$. That is, the interlinking capacitors are perturbative with respect to the on-site capacitances. Using the usual circuit-quantization techniques, we can write down the Lagrangian~\cite{Yurke1984,vool2017}:

\begin{equation}
    \mathcal{L}=\sum_{n=0}^N\frac{C_n\dot{\phi}_n^2}{2}
    -\frac{\phi_n^2}{2L_n}+\sum_{p\in B_n}\frac{C_{n,p}}{4}(\dot{\phi}_{p}-\dot{\phi}_n)^2
\end{equation}
where $\phi_n$ is the nodal flux at site $n$ and $B_n$ denotes the set of all site indices connected to site $n$. The extra factor of $1/2$ on the second term is to account for the double counting. We can write down the conjugate variables $q_n=\partial\mathcal{L}/\partial\dot{\phi}_n$:

\begin{equation}\label{eqn-appen-qn}
	q_n
	=C_{n(B)}\dot{\phi}_n - \sum_{p\in B_n}C_{n,p}\dot{\phi}_p,
\end{equation}
where $C_{n(B)}$ is the sum total of all capacitances connected to site $n$. Now by the Legendre's transformation $\mathcal{H}=\sum_{n=0}^Nq_n\dot{\phi}_n-\mathcal{L}$:

\begin{equation}
    \mathcal{H}=\sum_{n=0}^N\frac{C_{n(B)}}{2}\dot{\phi}^2_n+\frac{\phi_n^2}{2L_n}
    -\sum_{n\in B_n}\frac{C_{n,p}}{2}\dot{\phi}_n\dot{\phi}_{p}.
\end{equation}
Now to insert $q_n$ into this equation, we have to invert Eqn.~\ref{eqn-appen-qn}. If one treats it as a system of equations where $\mathbf{q}=(\mathbf{D}+\mathbf{X})\dot{\mathbf{\phi}}$ where $\mathbf{q}$ is a vector of $q_n$, $\dot{\mathbf{\phi}}$ is a vector of $\dot{\phi}_n$, $\mathbf{D}$ is a diagonal matrix with entries $C_{n(B)}$ and $\mathbf{X}$ is an off-diagonal matrix with entries $-C_{n,p}$. To invert $(\mathbf{D}+\mathbf{X})$, take $C_{n,p}$ to be perturbative compared to $C_{n(B)}$, to enable the approximation: $(\mathbf{D}+\mathbf{X})^{-1}= \mathbf{D}^{-1}-\mathbf{D}^{-1}\mathbf{X}\mathbf{D}^{-1}+\mathcal{O}(\|\mathbf{X}\|^2)$. From this one may show that $(\mathbf{D}^{-1}\mathbf{X}\mathbf{D}^{-1})_{ij}=-C_{i,j}/(C_{i(B)}C_{j(B)})$ to get:

\begin{equation}
    \dot{\phi}_n\approx\frac{q_n}{C_{n(B)}}+\sum_{p\in B_n}\frac{C_{n,p}}{C_{n(B)}C_{p(B)}}q_p.
\end{equation}
Note that this approximation simply removes higher-order terms that form multi-photon terms in the final expressions. This approximation holds as long as the operating powers and temperatures are low enough to restrict the system to single photon excitations. Substituting this expression into the Hamiltonian while discarding terms on the order $\mathcal{O}((C_{n,p}/C_{n(B)})^2)$ yields:

\begin{equation}
    \mathcal{H}\approx\sum_{n=0}^N\frac{q_n^2}{2C_{n(B)}}+\frac{\phi_n^2}{2L_n}
    +\sum_{p\in B_n}^{N}\frac{C_{n,p}}{2C_{n(B)}C_{p(B)}}q_nq_p.
\end{equation}
Now recognise the canonical harmonic oscillator form to get:

\begin{align}
    \omega_n&=\frac{1}{\sqrt{L_nC_{n(B)}}}\\
    Z_n&=\sqrt{\frac{C_{n(B)}}{L_n}}\\
    \hat{a}^\dagger&=\frac{1}{\sqrt{2\hbar Z_n}}(Z_n\phi_n-iq_n)\\
    \hat{a}&=\frac{1}{\sqrt{2\hbar Z_n}}(Z_n\phi_n+iq_n).
\end{align}
Now noting that $q_n=\frac{i}{2}\sqrt{2\hbar Z_n}(\hat{a}^\dagger_n-\hat{a}_n)$, the Hamiltonian simplifies into:

\begin{align}
    \mathcal{H}\approx&\sum_{n=0}^N\hbar\omega_n(\hat{a}^\dagger\hat{a}+\tfrac{1}{2})\nonumber\\
    &-\frac{\hbar}{2}\sum_{n=1}^{N}\sum_{m\in B_n}\frac{C_{n,m}\sqrt{Z_nZ_m}}{2C_{n(B)}C_{m(B)}}(\hat{a}^\dagger_n\hat{a}_m+\hat{a}^\dagger_m\hat{a}_n).
\end{align}
Now per the usual rotating wave approximation,  $(\hat{a}^\dagger_n-\hat{a}_n)(\hat{a}^\dagger_m-\hat{a}_m)\approx\hat{a}^\dagger_n\hat{a}_m+\hat{a}^\dagger_m\hat{a}_n$. The approximation works as the $\hbar\omega_n$ term dominates the tunnel-coupling term (noting that we can always add a constant offset to all sites to achieve this requirement). Now if one compares this Hamiltonian with the usual site model, this yields the relations for the on-site potentials and tunnel-couplings shown in the main text (note the double counting when swapping $m$ and $n$).

To include the effects of a lead connected to site $n$, we use convert the transmission line into a resistance $R$ (typically taken as $50\,\Omega$) connected via a coupling capacitor $c_{n}$. When taken over the limit $c_{n}\ll C_n$, the photon loss-rate on the resonator $L_n$-$C_n$ into the transmission line is~\cite{Eichler2013}:

\begin{equation}
    \kappa_n\approx\frac{c_n^2R}{C_n^2L_n}
\end{equation}
with a frequency shift $\Delta\omega_n$ giving:

\begin{equation}
    \frac{\Delta\omega_n}{\omega_n}=\frac{1}{\sqrt{1+\frac{c_n}{C_n}}}-1.
\end{equation}
The photon loss rate and frequency shift will correspond to the perturbations to the imaginary and real parts of the corresponding energy eigenvalue~\cite{mesoDatta}. This is equivalent to the self-energy term added to that site in the Hamiltonian. The proof is self-evident when considering the fact that the self-energy matrix $\boldsymbol{\Sigma}_n$ is all zero with the on-diagonal component $n$ being non-zero. Thus, $\boldsymbol{\Sigma}_n\mathbf{H}$ is symmetric (and noting that the individual matrices are also symmetric) to imply that $\boldsymbol{\Sigma}_n$ and $\mathbf{H}$ commute. This implies that $\boldsymbol{\Sigma}_n$ and $\mathbf{H}$ are simultaneously diagonalisable to realise that the net eigenvalues of $\mathbf{H}+\boldsymbol{\Sigma}_n$ are simply the summation. Therefore, noting that the eigenvalue $E_n$ satisfies the lifetime given as $-2\textnormal{Im}(E_n)$ with the frequency shift $\textnormal{Re}(E_n)\leqslant0$, the appropriate self-energy term to add onto the diagonal entry $n$ is thus:

\begin{equation}
    \Sigma_n=\Delta\omega_n-i\frac{\kappa_n}{2},
\end{equation}
to give:

\begin{equation}
    \Gamma_n=i(\Sigma_n-\Sigma_n^*)=\kappa_n.
\end{equation}

\section{Verifying cQED site model with SPICE simulations}\label{appen-cQED-SPICE}

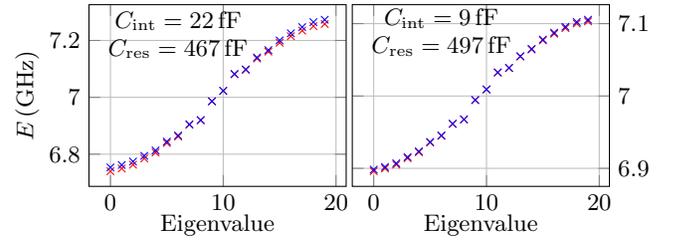
\begin{figure}[!ht]
	\input{FigNumPeaks}
	\caption{\label{fig:NumPeaks} Numerical simulations comparing the ideal eigenenergies with SPICE simulations of the cQED circuit implementation. We take $p=4$ cells in the Rice-Mele chains. The blue crosses represent ideal eigenvalues while red crosses represent energies in the corresponding cQED implementation (taken from peak positions in a transmission measurement through both ends of the Rice-Mele chain using $10\,\textnormal{G}\Omega$ input impedances). Both plots take a global energy offset of $7\,\textnormal{GHz}$ with $V_\textnormal{Q}=0$ and a constant inductance on every site of $1\,\textnormal{nH}$. The left plot takes $V_\textnormal{L/R}=\pm37.5\,\textnormal{MHz}$, $t_1=110\,\textnormal{MHz}$, $t_2=150\,\textnormal{MHz}$ and $t_\textnormal{Q}=75\,\textnormal{MHz}$. The right plot takes $V_\textnormal{L/R}=\pm15\,\textnormal{MHz}$, $t_1=60\,\textnormal{MHz}$, $t_2=48\,\textnormal{MHz}$ and $t_\textnormal{Q}=30\,\textnormal{MHz}$. The plot labels show the maximum interlinking capacitance $C_\textnormal{int}$ and minimum resonator capacitance $C_\textnormal{res}$. A clear feature is that when taking the limit $C_\textnormal{int}\ll C_\textnormal{res}$, the energy eigenvalues of the cQED implementation converge to the true ideal eigenvalues.}
\end{figure}

To verify that these parameters can be realised in a realistic experiment, some simple numeric simulations were performed. Here we compared the true eigenvalues, when numerically diagonalising the Hamiltonian matrix, with SPICE simulations of the translated cQED circuit when using the discussed recipe. The energy eigenvalues were taken from the SPICE simulation by taking the positions of the transmission peaks on probing sites $1$ and $19$ with $10\,\textnormal{G}\Omega$ impedances. The qubit site in the SPICE is treated as an LC oscillator with its energy set to $V_\textnormal{Q}=0$. We nominally centre the spectrum at $7\,\textnormal{GHz}$ and set the inductors to $1\,\textnormal{nH}$. When taking on-site potentials and tunnel couplings in the order of $10\,\textnormal{MHz}$ and $100\,\textnormal{MHz}$, we get resonator and interlinking capacitances around $22\,\textnormal{fF}$ and $467\,\textnormal{fF}$. These values are all experimentally feasible and can be fabricated well within the lumped element approximation. We see that when we use smaller tunnel couplings, to lower the value of the interlinking capacitances to around $9\,\textnormal{fF}$, the peaks in the cQED implementation better correspond to the ideal Hamiltonian as we have better satisfied the limit $C_{m,n}\ll C_{n(B)}$ required for a faithful implementation.

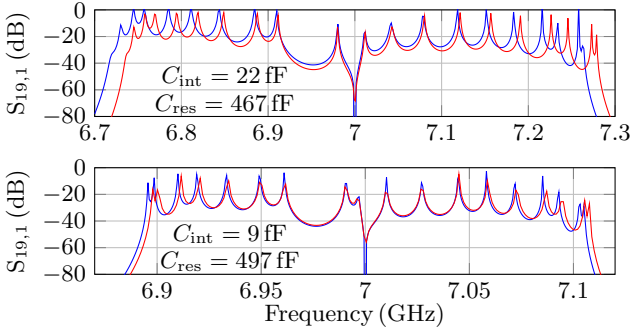
\begin{figure}[!ht]
	\input{FigNumZgreen}
	\caption{\label{fig:NumZgreen} Numerical simulations of the $S_{19,1}$ parameter when implementing in a cQED structure as measured through either end of the Rice-Mele chain. The blue curves represent ideal $S_{19,1}$ while red curves represent the corresponding cQED implementation (taken using $50\,\Omega$ input impedances coupled via $20\,\textnormal{fF}$ input capacitances). The remaining parameters for the two plots are the same as the corresponding plots shown in Fig.~\ref{fig:NumPeaks}. The plot labels show the maximum interlinking capacitance $C_\textnormal{int}$ and minimum resonator capacitance $C_\textnormal{res}$. A clear feature is that when taking the limit $C_\textnormal{int}\ll C_\textnormal{res}$, the transmission peaks of the cQED implementation converge to the true ideal peaks.}
\end{figure}

We also verify the expressions for the port couplings. Fig.~\ref{fig:NumZgreen} shows the transmission from site 1 to site 19 via $50\,\Omega$ ports coupled via $20\,\textnormal{fF}$ capacitors. The two plots correspond to the parameters shown in the two plots from Fig.~\ref{fig:NumPeaks}. We observe once again that the correspondence between the ideal peaks and the cQED peaks peaks gets better on respecting the limits $C_{m,n}\ll C_{n(B)}$ and $c_n\ll C_n$.

Finally, it is important to check whether defects in the proposed circuits, using fixed lumped elements, will affect the chirality. We take modest estimates for the accuracy in $V$, $t_1$ and $t_2$ to be $1\%$ given the $10~\textnormal{nm}$ precision in creating $1~\upmu\textnormal{m}$ structures when using electron beam lithography. After running 10,000 randomised samples of the resulting port-coupled circuits shown in Figures \ref{fig:NumPeaks} and \ref{fig:NumZgreen}, we obtain an average left/right chirality of $\chi=1130\pm80$. In the case of no port-couplings (that is, coupling the chains to adjacent qubits like in Fig.~\ref{fig:FigNumIdeal}e), the median chirality bounded by the 5th and 95th quantiles is 48000 ($6\times10^3$,$6\times10^6$). That is, the chirality is mostly unaffected by local defects.

\section{Transmission and reflection in terms of Green's functions}\label{sec:appen:mainGreen}

\begin{figure}[!ht]
	\input{figAppenGreen}
	\caption{\label{fig:appenGreenMain} A general Y-chain model as probed by ports. The central site $M$ is tunnel-coupled to two Rice-Mele chains (via $t_L$ and $t_R$) each with $p$ pairs sites with $t_2$ tunnel coupling, with each pair linked via tunnel coupling $t_1$. The coupling to the ports is given as $\Gamma_1$ and $\Gamma_N$. The on-site potentials are alternated via $\pm V$ with the central and qubit sites set to $V_M$ and $V_Q$ respectively. Note that the sites are enumerated from $1$ to $N=2M-1$ along the chain with the qubit site indexed as $N+1$ (only the indices $1$, $M$ and $N$ are labelled for clarity).}
\end{figure}
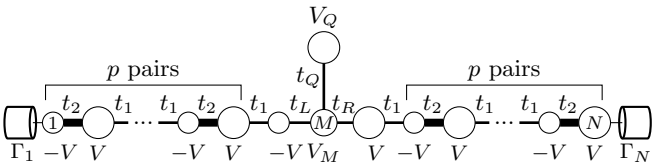

Consider a chain with $N=2M-1$ sites, with site $M$ side-coupled to the qubit $Q$ as shown in Fig.~\ref{fig:appenGreenMain}. We refer to sites from $1$ to $M-1$ as the left chain, and sites $M+1$ to $N$ the right chain. The central site $M$ is tunnel-coupled to site $Q=N+1$ via tunnel coupling $t_Q$. We consider the nearest neighbour hopping between adjacent sites, as well as various onsite energies on each site. In addition, we add the port-induced self-energy $-i \Gamma_1 \, | 1 \rangle \langle 1 | -  i \Gamma_N \, | N \rangle \langle N |$ to get the effective Hamiltonian $H$ of the whole model. Writing it in the block-matrix form
\begin{align}
H = \left( \begin{array}{cccc} V_Q & - t_Q & 0 & 0  \\ -t_Q & V_M & -T_L & -T_R \\ 0 & - T_L^{\dagger} & H_L & 0 \\ 0 & - T_R^{\dagger}  & 0 & H_R \end{array}\right) =  \left( \begin{array}{cc} H_c & -T \\ - T^{\dagger} & H_l \end{array} \right),
\end{align}
where $T_L = t_L | M\rangle \langle M -1|$ and $T_R = t_R | M\rangle \langle M +1|$, we compute the Green's function $G (\omega) = \frac{1}{\omega - H- \Sigma_l}$, where $\Sigma_l = \Sigma_L + \Sigma_R = - i \Gamma_1 |1 \rangle \langle 1 | - i \Gamma_N |N \rangle \langle N|$, by means of the formula
\begin{align}
 & \left( \begin{array}{cc} A & B \\ C & D \end{array}\right)^{-1} =  \left( \begin{array}{cc} L &  -  L B D^{-1} \\ - D^{-1} C  L  & D^{-1}+  D^{-1} C L B D^{-1} \end{array}\right) ,
\end{align}
where $L =  (A - B D^{-1} C)^{-1}$. Noting that all listed Green's functions are implicitly a function of $\omega$, for the remainder of this paper we drop the $\omega$ script for clarity. The overall Green's function for the above Hamiltonian is:

\begin{equation}
G (\omega) \label{GF_full} =  \left( \begin{array}{cc} G^c & - G^c T G^l \\ - G^l T^{\dagger} G^c &  G^l + G^l T^{\dagger} G^c T G^l  \end{array} \right) .
\nonumber
\end{equation}
In this expression we identify: 1) the tunneling terms $B=T$ and $C=T^{\dagger}$  out of and into the chains; 2) the Green's function of the decoupled chains:

\begin{align}
D^{-1} = G^l (\omega) = \frac{1}{\omega - H_l- \Sigma_l} = \left( \begin{array}{cc} G^L & 0 \\ 0 & G^R  
\end{array} \right) 
\label{GF_leads_bare}
\end{align}
where $G^{L,R} (\omega ) =(\omega - H_{L,R} - \Sigma_{L,R})^{-1}$; 3) the bare Green's function 
\begin{align}
A^{-1} = \bar{G}^c (\omega) = (\omega - H_c)^{-1}
\end{align}
of the isolated central subsystem $Q+M$;  4) its counterpart dressed by the coupling of the site $M$ to the chains 
\begin{align}
L &= G^c (\omega) = ([\bar{G}^c (\omega)]^{-1}  - T G^l (\omega) T^{\dagger})^{-1} \\
&= \left(  \begin{array}{cc} \omega - V_Q & t_Q \\ t_Q & \omega - V_M - \Sigma_M (\omega)  \end{array} \right)^{-1} \\
&=\frac{1}{d (\omega)} \left(  \begin{array}{cc} \omega - V_M - \Sigma_M (\omega) &  -t_Q \\ -t_Q &\omega - V_Q  \end{array}  \right),
\label{GF_central_dressed}
\end{align}
with 
\begin{align}
    \Sigma_M (\omega) = t_L^2 G^L_{M-1,M-1} (\omega) + t_R^2 G^R_{M+1,M+1} (\omega)
\end{align}
and
\begin{align}
d (\omega) = [\omega - V_Q] [\omega - V_M - \Sigma_M (\omega)]- t_Q^2 .
\end{align}

Transmission and reflection amplitudes are expressed by the formulas~\cite{mesoDatta}:

\begin{align}
S_{N1} (\omega) &= -2 i \sqrt{\Gamma_N \Gamma_1} \langle N | G (\omega) | 1 \rangle, \label{trans1} \\
S_{1N} (\omega) &= -2 i \sqrt{\Gamma_1 \Gamma_N} \langle 1 | G (\omega) | N \rangle, \label{trans2}
\end{align}
and
\begin{align}
S_{11} (\omega) &=1 - 2 i \Gamma_1  \langle 1 | G (\omega) | 1 \rangle, 
\label{refl1} \\
S_{NN} (\omega) &= 1- 2 i \Gamma_N  \langle N | G (\omega) | N \rangle,
\label{refl2}
\end{align}
respectively. On the basis of \eqref{GF_full} we evaluate
\begin{align}
 \langle N | G (\omega) | 1 \rangle &=  t_R t_L G^R_{N,M+1}  G^L_{M-1,1}   \frac{\omega - V_Q}{d (\omega)} , \\
\langle 1 | G (\omega) | N \rangle &=  t_L t_R G^L_{1,M-1}  G^R_{M+1,N}   \frac{\omega - V_Q}{d (\omega)} , 
\end{align}
and 
\begin{align}
 \langle 1 | G (\omega) | 1 \rangle &= G^L_{1,1}  +  t_L^2 G^L_{1,M-1}  G^L_{M-1,1}   \frac{\omega - V_Q}{d (\omega)} , \label{G11_app} \\
\langle N | G (\omega) | N \rangle &= G^{R}_{N,N}  + t_R^2 G^R_{N,M+1}  G^R_{M+1,N}   \frac{\omega - V_Q}{d (\omega)}. \label{GNN_app}
\end{align}
In turn, the port-dressed Green's functions of the left chain $G^L_{n,n'}$ ($n,n'=1, \ldots, M-1$) and of the right chain $G^R_{n,n'}$ ($n,n'=M+1, \ldots, N$) can be expressed in terms of the bare Green's functions $\bar{G}^L_{n,n'}$ and $\bar{G}^R_{n,n'}$ which describe the isolated left and right chains. Explicit formulas follow from the solution of the Dyson equations accounting the port-induced self-energy $\Sigma_l$ (see Appendix \ref{section:GFdetails:dressed} for details). Expressions for the relevant components, which occur in the Eqs.~\eqref{trans1}-\eqref{refl2},  read
\begin{align}
G^L_{1,1} &=   \frac{\bar{G}^L_{1,1}}{1+i  \Gamma_1 \bar{G}^L_{1,1}} , \\
G^L_{1,M-1} &=  G^L_{M-1,1}=  \frac{\bar{G}^L_{1,M-1}}{1+i  \Gamma_1 \bar{G}^L_{1,1}} , \\
G^L_{M-1,M-1} &= \frac{\bar{G}^L_{M-1,M-1}}{1+i  \Gamma_1 \bar{G}^L_{1,1}} \nonumber \\ 
&+i \Gamma_1   \frac{\bar{G}^L_{M-1,M-1} \bar{G}^L_{1,1} -\bar{G}^L_{M-1,1}\bar{G}^L_{1,M-1}}{1+i  \Gamma_1 \bar{G}^L_{1,1}} 
\end{align}
and
\begin{align}
G^R_{N,N} &=  \frac{\bar{G}^R_{N,N} }{1+i \Gamma_N \bar{G}^R_{N,N}}, \\
G^R_{N,M+1} &= G^R_{M+1,N}= \frac{\bar{G}^R_{N,M+1} }{1+i \Gamma_N \bar{G}^R_{N,N}}, \\
G^R_{M+1,M+1} &= \frac{\bar{G}^R_{M+1,M+1}}{1+i \Gamma_N \bar{G}^R_{N,N}}
\nonumber \\
&+ i \Gamma_N\frac{ \bar{G}^R_{M+1,M+1}\bar{G}^R_{N,N} -\bar{G}^R_{M+1,N} \bar{G}^R_{N,M+1} }{1+i \Gamma_N \bar{G}^R_{N,N}}.
\end{align}
Finally, choosing $N=4p+3 = 2 M-1$, where $p \geq 0$ is integer, we quote analytic expressions for the relevant ("far", "near", and "cross") components of the bare Green's functions of the chains (see Appendix \ref{section:GFdetails:bare})
\begin{align}
    \bar{G}^L_{1,1} &= \frac{w_f}{\omega + V}, \quad \bar{G}^R_{N,N} = \frac{w_f}{\omega -V}, \label{G_far} \\
     \bar{G}^L_{1,M-1} &= \bar{G}^L_{M-1,1} = \frac{w_c}{\omega + V}, \label{G_cross_l} \\ 
     \bar{G}^R_{N,M+1} &= \bar{G}^R_{M+1,N} = \frac{w_c}{\omega - V}, \label{G_cross_r} \\
     \bar{G}^L_{M-1,M-1} &= \frac{w_n}{\omega + V}, \quad \bar{G}^R_{M+1,M+1} = \frac{w_n}{\omega -V}, \label{G_near}
\end{align}
where 
\begin{align}
    w_f &= -\frac{F_0  (1  - Q^2) [  Q t_2 + t_1  -  Q^{2p+1} (t_2 + Q t_1)]}{2  Q (1-Q^{2p+2}) t_1}, \\
    w_n &=  -\frac{F_0 (1-Q^2) [t_2 + Q t_1 - Q^{2p+1} (Q t_2 + t_1)]}{2  Q (1-Q^{2p+2}) t_2}, \\
    w_c &=  -\frac{F_0}{2 Q} \frac{Q^{p} ( 1 - Q^2 )^2}{(1 -Q^{2p+2})} 
\end{align}
are frequency-dependent functions expressed via
\begin{align}
    F_0 &=  -\frac{1}{a\sqrt{1- \frac{1}{a^2}}}, \\
    Q &=  a + \frac{1}{F_0}, \\
    a &= \frac{\omega^2 - V^2 - t_1^2 - t_2^2}{2 t_1 t_2}.
\end{align}
Note that $F_0$ features branch-cuts in the complex frequency plane corresponding to the energy bands, and therefore it is meaningful to add a small positive imaginary term to frequency, $\omega \to \omega + i \eta$, to achieve the retarded Green's functions.

Finally, we obtain the following transmittance  formula
\begin{align}
& S_{N1} (\omega) = -  \frac{2 i \sqrt{\Gamma_N \Gamma_1}  t_R t_L  w_c^2}{(\omega - V+i \Gamma_R) (\omega+V+i  \Gamma_L)}  \frac{\omega - V_Q}{d (\omega)} ,
\end{align}
as well as the formulas for the reflectances
\begin{align}
& S_{11} (\omega) = 1 \nonumber \\
&-2i \Gamma_1 \left( \frac{w_f}{\omega+V+i  \Gamma_L} + \frac{t_L^2 w_c^2}{(\omega+V+i  \Gamma_L)^2}  \frac{\omega - V_Q}{d (\omega)} \right), \\
& S_{NN} (\omega) = 1 \nonumber \\
&-2i \Gamma_N \left( \frac{w_f}{\omega-V+i  \Gamma_R} + \frac{t_R^2 w_c^2}{(\omega-V+i  \Gamma_R)^2}  \frac{\omega - V_Q}{d (\omega)} \right). 
\end{align}
We discuss the implications of these equations in the main text.
Nonetheless, the cumbersome algebra can be simplified (making the insights clearer) via the approximations discussed in the following subsections.

\subsection{Effective 4-site model}

The discussion thus, far refer to exact analytic expressions. However, in the case of a large band gap, the states in the band gap can be approximated via the effective 4-site model discussed in the main text. 
That is, in the limit $\omega \to \pm V$ (note that this values lie in the band gap $[-\Delta, \Delta]$, $\Delta =\sqrt{(t_2 - t_1)^2+V^2}$), we find $Q = - \frac{t_1}{t_2} \equiv -e^{-\tfrac{1}{\xi}}$ and $F_0 = \frac{2 t_1 t_2}{t_2^2 - t_1^2} = \frac{2 e^{-\tfrac{1}{\xi}}}{1-e^{-\tfrac{2}{\xi}}}$ (assuming that $t_2 > t_1$ and hence
$\tfrac{1}{\xi} >0$). Hence
\begin{align}
w_f & \approx   \frac{e^{-2 p \tfrac{1}{\xi}}}{\mathcal{N}}  , \\
w_n & \approx \frac{1}{\mathcal{N}}  , \\
w_c  &\approx \frac{ (-1)^p e^{-p \tfrac{1}{\xi}}}{\mathcal{N}} ,
\end{align}
where
\begin{align}
    \mathcal{N} =  \frac{1 -   e^{-2( p+1) \tfrac{1}{\xi}}}{ 1 - e^{-2\tfrac{1}{\xi}}} .
\end{align}
These expressions are easily interpreted in terms of the edge state wavefunctions:
\begin{align}
    \bar{G}^L_{n.n'} & \approx \frac{\psi^{(e,L)}_n \psi^{(e,L)}_{n'}}{\omega + V}, \\ 
    \psi^{(e,L)}_{M-2|l|-1} &= \frac{(-1)^l e^{-\tfrac{1}{\xi} |l|}}{\sqrt{\mathcal{N}}}, \quad -p \leq l \leq 0 , \\
    \bar{G}^R_{n.n'} & \approx \frac{\psi^{(e,R)}_n \psi^{(e,R)}_{n'}}{\omega - V}, \\
    \psi^{(e,L)}_{M+2l+1} &= \frac{(-1)^l e^{-\tfrac{1}{\xi} l}}{\sqrt{\mathcal{N}}}, \quad 0 \leq l \leq p .
\end{align}
We also find that

\begin{align}
    \Delta w & \equiv w_f w_n - w_c^2 = \tilde{w} \frac{\omega^2 - V^2}{t_1 t_2}, \\
    \tilde{w} &= \left[ - \frac{F_0}{2 Q} (1-Q^2)\right]^2 \frac{Q (1 - Q^{2p}) }{ (1-Q^{2p+2}) }.
\end{align}
Collecting all contributions we express

\begin{align}
G^L_{1,1} &\approx   \frac{w_f}{\omega+V+i  \Gamma_L} , \\
G^L_{1,M-1} &\approx   \frac{w_c}{\omega+V+i  \Gamma_L} , \\
G^L_{M-1,M-1} &\approx \frac{w_n +i \Gamma_1 \tilde{w} \frac{\omega -V}{t_1 t_2}}{\omega +V+i  \Gamma_L} .
\end{align}
and
\begin{align}
G^R_{N,N} &\approx  \frac{w_f }{\omega - V+i \Gamma_R}, \\
G^R_{N,M+1} &\approx  \frac{w_c }{\omega - V+i \Gamma_R}, \\
G^R_{M+1,M+1} &\approx \frac{w_n + i \Gamma_1 \tilde{w} \frac{\omega+V}{t_1 t_2} }{\omega - V +i \Gamma_R}.
\end{align}
where $\Gamma_L = \Gamma_1 w_f$ and $\Gamma_R = \Gamma_N w_f$.

\subsection{Effective 3-site model}

We can continue the approximation even further by setting the on-site potential on central site $M$ to be very large. This effectively leaves 3 states in the band-gap to yield the effective 3-site model that excludes the central site. For large $V_M$ we approximate

\begin{align}
    d (\omega) & \approx - V_M \left( \omega - \tilde{V}_Q + i \gamma (\omega) \right), \\
    \tilde{V}_Q & = V_Q  - \frac{t_Q^2}{V_M}, \\
    \gamma (\omega) & = \frac{t_Q^2}{V_M^2} \left( \frac{t_L^2 \Gamma_L}{(\omega + V)^2 + \Gamma_L^2} + \frac{t_R^2 \Gamma_R}{(\omega - V)^2 + \Gamma_R^2} \right).
\end{align}
Hence
\begin{align}
& S_{11} (\omega) \approx 1 -\frac{2i \Gamma_L}{\omega+V+i  \Gamma_L} \nonumber \\
& \times \left( 1 - \frac{t_L^2}{\mathcal{N} V_M (\omega+V+i  \Gamma_L)}  \frac{\omega - V_Q}{\omega - \tilde{V}_Q + i \gamma (\omega)} \right), \\
& S_{NN} (\omega) = 1 -\frac{2i \Gamma_R}{\omega-V+i  \Gamma_R}  \nonumber \\
&\times \left( 1 - \frac{t_R^2}{\mathcal{N}  V_M(\omega-V+i  \Gamma_R)}  \frac{\omega - V_Q}{\omega - \tilde{V}_Q+i \gamma (\omega)} \right). 
\end{align}
Assuming that $V \gg \Gamma_{L,R}$ we also note that
\begin{align}
    \lim_{\omega \to \mp V} \gamma (\omega) \approx \frac{t_Q^2}{V_M^2} \frac{t_{L,R}^2}{\Gamma_{L,R}} \equiv{\gamma}_{L,R}.
\end{align}

\section{Details of the Green's function calculation}
\label{section:GFdetails}

\subsection{Port-dressed Green's functions of the chains}
\label{section:GFdetails:dressed}

The ports' self-energy in the site basis has the form
\begin{align}
    \Sigma_{n,n'}^L &= -i \Gamma_1 \delta_{n,1} \delta_{1,n'} , \\
    \Sigma_{n,n'}^R &= -i \Gamma_N \delta_{n,N} \delta_{N,n'} .
\end{align}

To compute the port-dressed Green's functions $G_{n,n'}^L$ ($n,n'=1, \ldots M-1$) and $G_{n,n'}^R$ ($n,n'=M+1, \ldots, N$) of the left and right chains, respectively, we solve the following Dyson equations
\begin{align}
G^L_{n,n'} &= \bar{G}^L_{n,n'} -i \Gamma_1 \bar{G}^L_{n,1} G^L_{1,n'}, 
\label{DysonL} \\
G^R_{n,n'} &= \bar{G}^R_{n,n'} -i \Gamma_N \bar{G}^R_{n,N} G^R_{N,n'}.
\label{DysonR}
\end{align}

Finding first the components
\begin{align}
G^L_{1,n'} &= \frac{\bar{G}^L_{1,n'}}{1+i  \Gamma_1 \bar{G}^L_{1,1}} , \\
G^R_{N,n'} &= \frac{\bar{G}^R_{N,n'} }{1+i \Gamma_N \bar{G}^R_{N,N}},
\end{align}
we insert these results into the initial equations \eqref{DysonL} and \eqref{DysonR} to obtain
\begin{align}
G^L_{n,n'} &= \bar{G}^L_{n,n'} -i \Gamma_1 \bar{G}^L_{n,1}  \frac{\bar{G}^L_{1,n'}}{1+i  \Gamma_1 \bar{G}^L_{1,1}} , \label{GFL_dressed} \\
G^R_{n,n'} &= \bar{G}^R_{n,n'} -i \Gamma_N \bar{G}^R_{n,N} \frac{\bar{G}^R_{N,n'} }{1+i \Gamma_N \bar{G}^R_{N,N}}.
\label{GFR_dressed}
\end{align}

\subsection{Evaluation of the bare Green's function of the isolated finite chains}
\label{section:GFdetails:bare}

In order to evaluate the Green's function $\bar{G}^L_{n,n'} $ of the finite right chain spanning over the sites from $M+1$ to $N$, we exploit the usual method to calculate the boundary Green's functions~\cite{PhysRevB.104.125447}. Using the Green's functions of the translationally invariant infinite chain as an unperturbed basis, we add the potential impurities to the sites $M$ and $N+1$. Solving the corresponding Dyson equation, one can exactly find the Green's function of the perturbed system. Finally, taking the strength of the impurity potentials to infinity one separates (in the case of solely nearest-neighbor hoppings) the finite chain region of interest from all the other sites. The solution of the Dyson equation discussed above reads in this limit 
\begin{widetext}
\begin{align}
\bar{G}_{n,n'}^R &=\bar{G}_{n-M,n'-M}^{(0,R)}  - \frac{1}{\bar{G}_{0,0}^{(0,R)} \bar{G}_{N-M+1,N-M+1}^{(0,R)}  -   \bar{G}_{0,N-M+1}^{(0,R)}  \bar{G}_{N-M+1,0}^{(0,R)} } \nonumber \\
&\times   \left[ \bar{G}_{N-M+1,N-M+1}^{(0,R)} \bar{G}_{n-M,0}^{(0,R)}  \bar{G}_{0,n'-M}^{(0,R)} +\bar{G}_{0,0}^{(0,R)}  \bar{G}_{n-M,N-M+1}^{(0,R)} \bar{G}_{N-M+1,n'-M}^{(0,R)}  \right. \nonumber \\
& \left. -  \bar{G}_{0,N-M+1}^{(0,R)}  \bar{G}_{n-M,0}^{(0,R)}  \bar{G}_{N-M+1,n'-M}^{(0,R)} - \bar{G}_{N-M+1,0}^{(0,R)} \bar{G}_{n-M,N-M+1}^{(0,R)}  \bar{G}_{0,n'-M}^{(0,R)} \right].
\end{align}
\end{widetext}
Hereby $\bar{G}_{n,n'}^{(0,R)}=\bar{G}_{n+2l,n' +2l}^{(0,R)}$ are the translationally invariant bulk Green's functions of the infinite chain. They are defined in terms of the Bloch Hamiltonian
\begin{align}
    h_k = \left( \begin{array}{cc} V & - t_1 - t_2 e^{-i k} \\
                                - t_1 - t_2 e^{ik} & -V 
                                \end{array} \right),
\end{align}
which corresponds to the choice of the unit cell $(V) \stackrel{-t_1}{\longleftrightarrow} (-V)$, with $-t_2$ being the intercell hopping amplitude. More explicitly, defining $\bar{G}_k^{(0,R)}= \frac{1}{\omega -h_k}$ we find the spatial components by means of the Fourier transformation. Performing analytically the integrals over the Bloch quasimomentum $k$, we obtain
\begin{align}
& \bar{G}^{(0,R)}_{2 l-1,2 l' -1} (\omega) = \int_{-\pi}^{\pi} \frac{d k}{2 \pi} e^{i k (l -l')}  [\bar{G}_k^{(0,R)} (\omega)]_{11} \nonumber \\
&=- \frac{\omega+V}{2 t_1 t_2} F_{0}  Q^{|l-l'|}, \\
& \bar{G}^{(0,R)}_{2 l,2 l' } (\omega) = \int_{-\pi}^{\pi} \frac{d k}{2 \pi} e^{i k (l -l')}  [\bar{G}_k^{(0,R)} (\omega)]_{22} \nonumber \\
&=- \frac{\omega-V}{2 t_1 t_2} F_{0} Q^{|l-l'|}, \\
\end{align}
\begin{align}
& \bar{G}^{(0,R)}_{2 l-1,2 l' } (\omega) = \int_{-\pi}^{\pi} \frac{d k}{2 \pi} e^{i k (l -l')}  [\bar{G}^{(0,R)}_k (\omega)]_{12} \nonumber \\
&= \frac{ F_{0} Q^{|l-l'|} }{2 t_1 t_2}  \left\{ \Theta_{l \leq l'} \left[ t_1 + t_2 Q  \right]  + \Theta_{l \geq l' +1}   \left[ t_1  + \frac{t_2}{Q} \right] \right\},  \\
& \bar{G}^{(0,R)}_{2 l,2 l' -1} (\omega) = \int_{-\pi}^{\pi} \frac{d k}{2 \pi} e^{i k (l -l')}  [\bar{G}^{(0,R)}_k (\omega)]_{21}\nonumber  \\
&= \bar{G}^{(0,R)}_{2 l' -1,2l} (\omega)  .
\end{align}

Analogously we find the left-chain Green's function
\begin{widetext}
\begin{align}
\bar{G}^L_{n,n'}&=\bar{G}_{n-M,n'-M}^{(0,L)}  - \frac{1}{\bar{G}_{-M,-M}^{(0,L)} \bar{G}_{0,0}^{(0,L)}  -   \bar{G}_{-M,0}^{(0,L)}  \bar{G}_{0,-M}^{(0,L)} } \nonumber \\
&\times \left[  \bar{G}_{0,0}^{(0,L)} \bar{G}_{n-M,-M}^{(0,L)}  \bar{G}_{-M,n'-M}^{(0,L)} +\bar{G}_{-M,-M}^{(0,L)}  \bar{G}_{n-M,0}^{(0,L)} \bar{G}_{0,n'-M}^{(0,L)}  \right. 
\nonumber \\
& \left. -  \bar{G}_{-M,0}^{(0,L)}  \bar{G}_{n-M,-M}^{(0,L)}  \bar{G}_{0,n'-M}^{(0,L)} - \bar{G}_{0,-M}^{(0,L)} \bar{G}_{n-M,0}^{(0,L)}  \bar{G}_{-M,n'-M}^{(0,L)} \right]
\end{align}
\end{widetext}
in terms of the translationally invariant bulk Green's functions $\bar{G}_{n,n'}^{(0,L)} = \bar{G}_{n+2l,n'+2l}^{(0,L)}$ corresponding to the unit cell choice $(-V) \stackrel{-t_2}{\longleftrightarrow} (V)$, with $-t_1$ being the intercell hopping amplitude.

Choosing even $M=2p+2$ (and thus $N=4p+3$) and using the identities $Q \bar{G}_{1,1}^{(0,R)} \bar{G}_{0,0}^{(0,R)} = \bar{G}_{-1,0}^{(0,R)} \bar{G}_{1,0}^{(0,R)}$ and $Q \bar{G}_{1,1}^{(0,L)} \bar{G}_{0,0}^{(0,L)} = \bar{G}_{-1,0}^{(0,L)} \bar{G}_{1,0}^{(0,L)}$, we evaluate the interesting components for the right chain
\begin{align}
\bar{G}_{N,N}^R &=\frac{[ \bar{G}_{1,0}^{(0,R)} - Q \bar{G}_{-1,0}^{(0,R)}] [\bar{G}_{-1,0}^{(0,R)}   -  Q^{2p+1} \bar{G}_{1,0}^{(0,R)}]}{Q \bar{G}_{0,0}^{(0,R)}(1-Q^{2 p +2}) } 
\end{align}
\begin{align}
\bar{G}_{M+1,M+1}^R &= \frac{[ \bar{G}_{-1,0}^{(0,R)} -  Q \bar{G}_{1,0}^{(0,R)} ] [\bar{G}_{1,0}^{(0,R)}-Q^{2 p +1} \bar{G}_{-1,0}^{(0,R)} ] }{Q \bar{G}_{0,0}^{(0,R)}  (1-Q^{2 p +2})} ,
\end{align}
\begin{align}
\bar{G}_{N,M+1}^R &=   \frac{Q^{p-1} [\bar{G}_{1,0}^{(0,R)} -  Q\bar{G}_{-1,0}^{(0,R)}] [\bar{G}_{-1,0}^{(0,R)}-Q \bar{G}_{1,0}^{(0,R)} ] }{\bar{G}_{0,0}^{(0,R)}  (1-Q^{2 p +2})  }  ,
\end{align}
and for the left chain
\begin{align}
\bar{G}^L_{1,1} &=\frac{[\bar{G}_{-1,0}^{(0,L)} - Q \bar{G}_{1,0}^{(0,L)}] [    \bar{G}_{1,0}^{(0,L)}  -  Q^{2p+1} \bar{G}_{-1,0}^{(0,L)} ] }{Q \bar{G}_{0,0}^{(0,L)}  (1-Q^{2p+2}) } ,
\end{align}
\begin{align}
& \bar{G}^L_{M-1,M-1}=\frac{ [\bar{G}_{1,0}^{(0,L)} -Q \bar{G}_{-1,0}^{(0,L)}] [ \bar{G}_{-1,0}^{(0,L)}   -  Q^{2p+1} \bar{G}_{1,0}^{(0,L)}  ]}{ Q \bar{G}_{0,0}^{(0,L)}  (1- Q^{2p+2})} ,
\end{align}
\begin{align}
 \bar{G}^L_{1,M-1} & =\frac{Q^{p-1}  [   \bar{G}_{1,0}^{(0,L)} -Q  G_{-1,0}^{(0,L)}  ] [    \bar{G}_{-1,0}^{(0,L)}  -Q \bar{G}_{1,0}^{(0,L)} ]}{\bar{G}_{0,0}^{(0,L)}  (1 -   Q^{2p+2}) } .
\end{align}
Inserting
\begin{align}
    \bar{G}_{1,0}^{(0,R)} = \bar{G}_{-1,0}^{(0,L)} = \frac{F_0}{2 t_1 t_2} (t_2 + Q t_1), \\
    \bar{G}_{-1,0}^{(0,R)} = \bar{G}_{1,0}^{(0,L)} =\frac{F_0}{2 t_1 t_2} (t_1 + Q t_2)
\end{align}
into the above expressions, we derive \eqref{G_far}-\eqref{G_near}.

\section{Local density of states}\label{appen:LDOS}

Noticing that the real parts of reflectances \eqref{refl1} and \eqref{refl1} provide us with the information about the local density of states (LDOS) $\rho_n = -\frac{1}{\pi}\,  \text{Im} \, G_{n,n}$ at the very first ($\rho_1 = \text{Re} \frac{1-S_{11}}{2 \pi \Gamma_1}$) and the very last ($\rho_N =\text{Re} \frac{1-S_{NN}}{2 \pi \Gamma_N}$) sites, we complement this discussion by presenting explicit formulas for $\rho_n$ at all the intermediate sites.

By the analogy with \eqref{G11_app} and  \eqref{GNN_app} we write down more general expressions for $1 \leq n \leq M-1$
\begin{align}
 \langle n | G (\omega) | n \rangle &= G^L_{n,n}  +  t_L^2 G^L_{n,M-1}  G^L_{M-1,n}   \frac{\omega - V_Q}{d (\omega)} ,
 \end{align}
 and for  $M+1 \leq n \leq N$
 \begin{align}
\langle n | G (\omega) | n \rangle &= G^{R}_{n,n}  + t_R^2 G^R_{n,M+1}  G^R_{M+1,n}   \frac{\omega - V_Q}{d (\omega)}.
\end{align}

\bibliography{apssamp}

\end{document}

%% file: FigNumIdeal2.tex
\newcommand{\miniFour}[4]{
	\node[draw, circ=5pt, fill=#4] (b) at (1.2,0.7) {};
	\node[draw, circ=5pt, fill=#3] (q) at (1.2,1.1) {};
	\draw[-, line width=0.5pt] (b) -- (q) node [pos=0.5, right]{};
	\node[draw, circ=7pt, fill=#2] (a) at (1.6,0.7) {};
	\coordinate[] (c) at (1.9,0.7) {} {};
	\draw[-, line width=0.5pt] (c) -- (a) node [pos=0.5, right]{};
	\draw[-, line width=0.5pt] (b) -- (a) node [pos=0.5, right]{};
	\node[draw, circ=3pt, fill=#1] (a) at (0.8,0.7) {};
	\coordinate[] (c) at (0.6,0.7) {} {};
	\draw[-, line width=0.5pt] (c) -- (a) node [pos=0.5, right]{};
	\draw[-, line width=0.5pt] (b) -- (a) node [pos=0.5, right]{};
}

\begin{tikzpicture}[
	circ/.style = {circle, minimum size=#1,
		inner sep=0pt, outer sep=0pt,
		even odd rule}
	]
	\tikzset{styleVplus/.style={inner color={rgb,255:red,255; green,128; blue,128},outer color=red, circ="10pt"}};
	\tikzset{styleVminus/.style={inner color={rgb,255:red,128; green,128; blue,255},outer color=blue, circ="6pt"}};
	\tikzset{styleTcBig/.style={line width="3pt"}};
	\tikzset{styleTcSmall/.style={line width="1pt"}};
	
	\begin{scope}[xshift=0.0cm, yshift=1.1cm]
		\node[](c) at (-3.8, 1.4) {\textbf{(a)}};
		
		\node[draw, circ=8pt] (b) at (0, 0) {};
		\node[draw, circ=8pt] (q) at (0, 0.9) {=};
		\node[right] at (q.east) {\footnotesize{$V_\textnormal{Q}$}};
		\draw[-, line width=0.5pt] (b) -- (q) node [pos=0.5, right]{\footnotesize{$t_\textnormal{Q}$}};
		\foreach \i in {1,...,4}
		{
			\node[draw, styleVminus] (a) at (-1.0*\i+0.5, 0) {};
			\ifnum \i=1
			\draw[-, styleTcSmall] (b) -- (a);
			\else
			\draw[-, styleTcBig] (b) -- (a) node [pos=0.5, above] {\ifnum\i=2 \footnotesize{$t_\textnormal{2}$} \fi};
			\fi
			\ifnum \i<4
			\node[draw, styleVplus] (b) at (-1.0*\i, 0) {};
			\draw[-, styleTcSmall] (b) -- (a);
			\fi
		}
		\draw[-, line width="1pt"] (a) -- ($ (a) - (0.25,0) $);
		\node[align=center] (a) at ($ (a) - (0.4,0) $) {...};
		
		\node[draw, circ=8pt] (b) at (0, 0) {};
		\foreach \i in {1,...,4}
		{
			\node[draw, styleVplus] (a) at (1.0*\i-0.5, 0) {};
			\ifnum \i=1
			\draw[-, styleTcSmall] (b) -- (a);
			\else
			\draw[-, styleTcBig] (b) -- (a);
			\fi
			\ifnum \i<4
			\node[draw, styleVminus] (b) at (1.0*\i, 0) {};
			\draw[-, line width="1pt"] (b) -- (a) node [pos=0.5, above] {\footnotesize{\ifnum\i=2 $t_\textnormal{1}$ \fi}};
			\fi
		}
		\draw[-, styleTcSmall] (a) -- ($ (a) + (0.25,0) $);
		\node[align=center] (a) at ($ (a) + (0.4,0) $) {...};
		\draw[draw=black, dashed] (-0.75,1.25) rectangle (0.75,-0.6);
		\node at (-0.5,-0.35) {L};
		\node at (0,0.-0.35) {C};
		\node at (0.5,-0.35) {R};
		\node at (-0.35,0.9) {Q};
		
		\draw[->] (-0.9,-0.3) -- (-0.9,-0.4) -- (-3.75,-0.4);
		\node at (-2.75,-0.6) {\footnotesize{$p$ unit cells}};
		\draw[->] (0.9,-0.3) -- (0.9,-0.4) -- (3.85,-0.4);
		\node at (2.85,-0.6) {\footnotesize{$p$ unit cells}};
		
		\draw[dashed]  (-2.75,0.3) rectangle (-1.75,-0.3);
		\node at (-2.25,0.5) {\footnotesize{unit cell}};
		
		\node at (2.9,0.3) {\footnotesize{\color{blue}$-V$}};
		\node at (3.5,0.3) {\footnotesize{\color{red}$+V$}};
		
		\node at (0,1.5) {\footnotesize{4 site model}};
	\end{scope}

\begin{scope}[xshift=-2.81cm, yshift=-5.5cm]
	\node[](c) at (-1, 5.3) {\textbf{(b)}};
		\begin{axis}[
		height=7cm,
		width=8.5cm,
		xmin=-4, xmax = 4,
		xlabel=$V_\textnormal{Q}$,
		ylabel=$E$,
		x label style={at={(axis description cs:0.5,-0.085)},anchor=south},
		y label style={at={(axis description cs:0.1,0.5)}, anchor=south},
		xtick={-2,0,2},
		xticklabels={$-V$,$0$,$V$},
		ytick={-7.07107,-2,0,2, 7.07107},
		yticklabels={$-\sqrt{2}t_1$, $-V$,$0$,$V$, $\sqrt{2}t_1$},
		grid
		]
		\addplot [black] table[mark=none, x index = {0},y index = {1}]{plotRM_Yconfig.tsv};
		\addplot [black] table[mark=none, x index = {0},y index = {2}]{plotRM_Yconfig.tsv};
		\addplot [black] table[mark=none, x index = {0},y index = {3}]{plotRM_Yconfig.tsv};
		\addplot [black] table[mark=none, x index = {0},y index = {4}]{plotRM_Yconfig.tsv};
		
		\addplot[blue, mark=o, mark size=3pt] coordinates {(-2.0,-2.0)};
		\addplot[blue, mark=x] coordinates {(-2.0,-2.0)};
		\addplot[red, mark=o, mark size=3pt] coordinates {(2.0,2.0)};
		\addplot[red, mark=+] coordinates {(2.0,2.0)};
	\end{axis}
	\node [] at (1.2,1.4) {$t_\textnormal{Q}\left|{\textnormal{L}}\right\rangle-t_1\left|{\textnormal{Q}}\right\rangle$};
	\node [] at (1.7,3) {$2\left|{\textnormal{R}}\right\rangle-\left|{\textnormal{L}}\right\rangle-\tfrac{V}{t_1}\left|{\textnormal{C}}\right\rangle$};
	\node [] at (5.7,4) {$t_\textnormal{Q}\left|{\textnormal{R}}\right\rangle-t_1\left|{\textnormal{Q}}\right\rangle$};
	\node [] at (5.3,2.4) {$2\left|{\textnormal{L}}\right\rangle-\left|{\textnormal{R}}\right\rangle+\tfrac{V}{t_1}\left|{\textnormal{C}}\right\rangle$};
	
	\draw [->] (3.45,2.3985) -- (3.45,2.6);
	\draw [->] (3.45,3.6) -- (3.45,2.8252);
	\node at (3.45,4) {$\dfrac{2V}{\sqrt{2}}\cdot\dfrac{t_\textnormal{Q}}{t_1}$};
	
	\begin{scope}
		\begin{axis}[at={(0.2cm,0.7cm)}, height=0.5cm, width=2cm, anchor=south west, scale only axis, hide axis, xmin=-4,xmax=4,ymin=0,ymax=1, xtick=\empty, ytick=\empty] 
			\addplot[shade, left color=white, right color=blue, draw=none, samples=100, domain=-4:0] {exp(-abs(0.5*x))} \closedcycle;
		\end{axis}
		\miniFour{white}{black}{white}{black};
	\end{scope}
	\begin{scope}[yshift=2.8cm]
		\begin{axis}[at={(0.2cm,0.7cm)}, height=0.3cm, width=2cm, anchor=south west, scale only axis, hide axis, xmin=-4,xmax=4,ymin=0,ymax=1, xtick=\empty, ytick=\empty] 
			\addplot[shade, left color=white, right color=blue, draw=none, samples=100, domain=-4:0] {exp(-abs(0.5*x))} \closedcycle;
		\end{axis}
		\begin{axis}[at={(0.2cm,0.7cm)}, height=0.3cm, width=2cm, anchor=south west, scale only axis, hide axis, xmin=-4,xmax=4,ymin=0,ymax=1, xtick=\empty, ytick=\empty] 
			\addplot[shade, left color=blue, right color=white, draw=none, samples=100, domain=0:4] {exp(-abs(0.5*x))} \closedcycle;
		\end{axis}
		\miniFour{white}{white}{black}{white};
	\end{scope}
	\begin{scope}[xshift=4.5cm, yshift=0.9cm]
		\begin{axis}[at={(0.2cm,0.7cm)}, height=0.3cm, width=2cm, anchor=south west, scale only axis, hide axis, xmin=-4,xmax=4,ymin=0,ymax=1, xtick=\empty, ytick=\empty] 
			\addplot[shade, left color=white, right color=red, draw=none, samples=100, domain=-4:0] {exp(-abs(0.5*x))} \closedcycle;
		\end{axis}
		\begin{axis}[at={(0.2cm,0.7cm)}, height=0.3cm, width=2cm, anchor=south west, scale only axis, hide axis, xmin=-4,xmax=4,ymin=0,ymax=1, xtick=\empty, ytick=\empty] 
			\addplot[shade, left color=red, right color=white, draw=none, samples=100, domain=0:4] {exp(-abs(0.5*x))} \closedcycle;
		\end{axis}
		\miniFour{white}{white}{black}{white};
	\end{scope}
	\begin{scope}[xshift=4.5cm, yshift=3.7cm]
		\begin{axis}[at={(0.2cm,0.7cm)}, height=0.5cm, width=2cm, anchor=south west, scale only axis, hide axis, xmin=-4,xmax=4,ymin=0,ymax=1, xtick=\empty, ytick=\empty] 
			\addplot[shade, left color=red, right color=white, draw=none, samples=100, domain=0:4] {exp(-abs(0.5*x))} \closedcycle;
		\end{axis}
		\miniFour{black}{white}{white}{black};
	\end{scope}
\end{scope}

	\begin{scope}[xshift=5.5cm, yshift=-1.2cm, spy using outlines={rectangle, connect spies}]
		\node[](c) at (-1, 3.7) {\textbf{(c)}};
		\begin{axis}[
			height=5.5cm,
			width=8.5cm,
			xlabel={Eigenvalue},
			ylabel={Energy~(MHz)},
			xmin=0, xmax=44,
			ymin=-300, ymax=300,
			x label style={at={(axis description cs:0.5,-0.13)}, anchor=south},
			y label style={at={(axis description cs:0.07,0.5)}, anchor=south},
			legend pos=south east,
			grid
			]
			\draw[draw=none,fill=black!20,opacity=0.5](axis cs: 0,-300)--(axis cs: 44,-300)--(axis cs: 44,-57.99)--(axis cs: 0,-57.99)--cycle;
			\draw[draw=none,fill=black!20,opacity=0.5](axis cs: 0,56.4)--(axis cs: 44,56.4)--(axis cs: 44,300)--(axis cs: 0,300)--cycle;
			\addplot[red, only marks, mark=+] table[x index = {0}, y index = {1}, col sep=comma] {idealSimPosEvals.csv};
			\addplot[blue, only marks, mark=x] table[x index = {0}, y index = {1}, col sep=comma] {idealSimNegEvals.csv};
			\legend{$V_\textnormal{Q}=+V$,$V_\textnormal{Q}=-V$};

			\addplot[blue, mark=o, mark size=3pt] coordinates {(22,-37.5)};
			\addplot[red, mark=o, mark size=3pt] coordinates {(23,37.5)};
			
			\node[align=center, text=black] at (axis cs:10,175) {Simulation\\for $p=10$};
		\end{axis}
	\end{scope}

	\begin{scope}[xshift=5.5cm, yshift=-5.5cm]
		\node[](c) at (-1, 2.8) {\textbf{(d)}};
		\begin{axis}[
			name=plot1,
			height=4.5cm,
			width=4.9cm,
			xlabel={Site Index},
			ylabel={Probability},
			xmin=0, xmax=44,
			ymin=0, ymax=0.19,
			yticklabel style={
				/pgf/number format/fixed,
				/pgf/number format/precision=5
			},
			x label style={at={(axis description cs:0.5,-0.13)}, anchor=south},
			y label style={at={(axis description cs:0.16,0.5)}, anchor=south},
			grid
			]
			\addplot[blue, fill=blue, fill opacity = 0.5] table[x index = {0}, y index = {1}, col sep=comma] {idealSimNegEvecs.csv};
			\node[align=center, text=blue] at (axis cs:22,0.17) {Left edge state $\bigotimes$};
			\node[align=center, text=black] at (axis cs:32,0.05) {No\\Leakage};
		\end{axis}
		\node[] (plt1) at (plot1.south east) {};
		\begin{axis}[
			at={($(plt1)+(0.2cm,0)$)},
			height=4.5cm,
			width=4.9cm,
			xlabel={Site Index},
			xmin=0, xmax=44,
			ymin=0, ymax=0.19,
			yticklabels={,,},
			x label style={at={(axis description cs:0.5,-0.13)}, anchor=south},
			y label style={at={(axis description cs:0.07,0.5)}, anchor=south},
			grid
			]
			\addplot[red, fill=red, fill opacity = 0.5] table[x index = {0}, y index = {2}, col sep=comma] {idealSimPosEvecs.csv};
			\node[align=center, text=red] at (axis cs:22,0.17) {Right edge state $\bigoplus$};;
			\node[align=center, text=black] at (axis cs:11,0.05) {No\\Leakage};
		\end{axis}
	\end{scope}

	\tikzset{styleVplus/.style={inner color={rgb,255:red,255; green,128; blue,128},outer color=red, circ="9pt"}};
	\tikzset{styleVminus/.style={inner color={rgb,255:red,128; green,128; blue,255},outer color=blue, circ="5pt"}};
	\tikzset{styleTcBig/.style={line width="3pt"}};
	\tikzset{styleTcSmall/.style={line width="1pt"}};
	\begin{scope}[xshift=0.0cm, yshift=-7.4cm]
		\node[](c) at (-3.8, 1.0) {\textbf{(e)}};
		\begin{scope}[xshift=-1.68cm]
			\begin{axis}[at={(-3cm,0.0cm)}, height=0.6cm, width=6cm, anchor=south west, scale only axis, hide axis, xmin=-4,xmax=4,ymin=0,ymax=1, xtick=\empty, ytick=\empty] 
				\addplot[fill=red, draw=none, samples=100, domain=0:4, opacity=0.5] {exp(-abs(0.4*x))} \closedcycle;
			\end{axis}
			\begin{axis}[at={(0.33cm,0.0cm)}, height=0.6cm, width=6cm, anchor=south west, scale only axis, hide axis, xmin=-4,xmax=4,ymin=0,ymax=1, xtick=\empty, ytick=\empty] 
				\addplot[fill=blue, draw=none, samples=100, domain=-4:0, opacity=0.5] {exp(-abs(0.4*x))} \closedcycle;
			\end{axis}
			
			\node[] at (1.665, 0.7) {\footnotesize{Finite interaction}};
			
			\node[draw, circ=8pt, fill=white] (b) at (0, 0) {};
			\node[draw, styleVplus] (q) at (0, 0.6) {};
			\node[right] at (q.east) {};
			\draw[-, line width=0.5pt] (b) -- (q) node [pos=0.5, right]{};
			\foreach \i in {1,...,3}
			{
				\node[draw, styleVminus] (a) at (-0.74*\i+0.37, 0) {};
				\ifnum \i=1
				\draw[-, styleTcSmall] (b) -- (a);
				\else
				\draw[-, styleTcBig] (b) -- (a) node [pos=0.5, above] {};
				\fi
				\ifnum \i<3
				\node[draw, styleVplus] (b) at (-0.74*\i, 0) {};
				\draw[-, styleTcSmall] (b) -- (a);
				\fi
			}
			\draw[-, line width="1pt"] (a) -- ($ (a) - (0.2,0) $);
			\node[align=center] (a) at ($ (a) - (0.37,0) $) {...};
			
			\node[draw, circ=8pt, fill=white] (b) at (0, 0) {};
			\foreach \i in {1,...,4}
			{
				\node[draw, styleVplus] (a) at (0.74*\i-0.37, 0) {};
				\ifnum \i=1
				\draw[-, styleTcSmall] (b) -- (a);
				\else
				\draw[-, styleTcBig] (b) -- (a);
				\fi
				\ifnum \i<4
				\node[draw, styleVminus] (b) at (0.74*\i, 0) {};
				\draw[-, line width="1pt"] (b) -- (a) node [pos=0.5, above] {};
				\fi
			}
			\node[draw, styleVminus] (b) at ($ (a) + (0.37,0) $) {};		
			\draw[-, styleTcSmall] (a) -- (b);
			\node[draw, circ=8pt, fill=white] (a) at ($ (b) + (0.37,0) $) {};
			\draw[-, styleTcSmall] (a) -- (b);
			\node[draw, styleVminus] (q) at ($ (a) + (0.0,0.6) $) {};
			\draw[-, styleTcSmall] (a) -- (q);
			\node[draw, styleVplus] (c) at ($ (a) + (0.37,0.0) $) {};
			\draw[-, styleTcSmall] (a) -- (c);
			\foreach \i in {1,...,3}
			{
				\node[draw, styleVminus] (a) at ($ (c) + (0.74*\i-0.37,0.0) $) {};
				\ifnum \i=1
				\draw[-, styleTcSmall] (c) -- (a);
				\else
				\draw[-, styleTcBig] (b) -- (a) node [pos=0.5, above] {};
				\fi
				\ifnum \i<3
				\node[draw, styleVplus] (b) at ($ (c) + (0.74*\i,0.0) $) {};
				\draw[-, styleTcSmall] (b) -- (a);
				\fi
			}
			\draw[-, line width="1pt"] (a) -- ($ (a) + (0.2,0) $);
			\node[align=center] (a) at ($ (a) + (0.37,0) $) {...};
		\end{scope}
		\begin{scope}[xshift=6.46cm]
			\begin{axis}[at={(-3cm,0.0cm)}, height=0.6cm, width=6cm, anchor=south west, scale only axis, hide axis, xmin=-4,xmax=4,ymin=0,ymax=1, xtick=\empty, ytick=\empty] 
				\addplot[fill=blue, draw=none, samples=100, domain=-2.75:0, opacity=0.5] {exp(-abs(0.4*x))} \closedcycle;
			\end{axis}
			\begin{axis}[at={(0.33cm,0.0cm)}, height=0.6cm, width=6cm, anchor=south west, scale only axis, hide axis, xmin=-4,xmax=4,ymin=0,ymax=1, xtick=\empty, ytick=\empty] 
				\addplot[fill=red, draw=none, samples=100, domain=0:3.25, opacity=0.5] {exp(-abs(0.4*x))} \closedcycle;
			\end{axis}
		
			\node[] at (1.665, 0.7) {\footnotesize{Zero interaction}};
			
			\node[draw, circ=8pt, fill=white] (b) at (0, 0) {};
			\node[draw, styleVminus] (q) at (0, 0.6) {};
			\node[right] at (q.east) {};
			\draw[-, line width=0.5pt] (b) -- (q) node [pos=0.5, right]{};
			\foreach \i in {1,...,3}
			{
				\node[draw, styleVminus] (a) at (-0.74*\i+0.37, 0) {};
				\ifnum \i=1
				\draw[-, styleTcSmall] (b) -- (a);
				\else
				\draw[-, styleTcBig] (b) -- (a) node [pos=0.5, above] {};
				\fi
				\ifnum \i<3
				\node[draw, styleVplus] (b) at (-0.74*\i, 0) {};
				\draw[-, styleTcSmall] (b) -- (a);
				\fi
			}
			\draw[-, line width="1pt"] (a) -- ($ (a) - (0.2,0) $);
			
			\node[draw, circ=8pt, fill=white] (b) at (0, 0) {};
			\foreach \i in {1,...,4}
			{
				\node[draw, styleVplus] (a) at (0.74*\i-0.37, 0) {};
				\ifnum \i=1
				\draw[-, styleTcSmall] (b) -- (a);
				\else
				\draw[-, styleTcBig] (b) -- (a);
				\fi
				\ifnum \i<4
				\node[draw, styleVminus] (b) at (0.74*\i, 0) {};
				\draw[-, line width="1pt"] (b) -- (a) node [pos=0.5, above] {};
				\fi
			}
			\node[draw, styleVminus] (b) at ($ (a) + (0.37,0) $) {};		
			\draw[-, styleTcSmall] (a) -- (b);
			\node[draw, circ=8pt, fill=white] (a) at ($ (b) + (0.37,0) $) {};
			\draw[-, styleTcSmall] (a) -- (b);
			\node[draw, styleVplus] (q) at ($ (a) + (0.0,0.6) $) {};
			\draw[-, styleTcSmall] (a) -- (q);
			\node[draw, styleVplus] (c) at ($ (a) + (0.37,0.0) $) {};
			\draw[-, styleTcSmall] (a) -- (c);
			\foreach \i in {1,...,3}
			{
				\node[draw, styleVminus] (a) at ($ (c) + (0.74*\i-0.37,0.0) $) {};
				\ifnum \i=1
				\draw[-, styleTcSmall] (c) -- (a);
				\else
				\draw[-, styleTcBig] (b) -- (a) node [pos=0.5, above] {};
				\fi
				\ifnum \i<3
				\node[draw, styleVplus] (b) at ($ (c) + (0.74*\i,0.0) $) {};
				\draw[-, styleTcSmall] (b) -- (a);
				\fi
			}
			\draw[-, line width="1pt"] (a) -- ($ (a) + (0.2,0) $);
			\node[align=center] (a) at ($ (a) + (0.37,0) $) {...};
		\end{scope}
	\end{scope}

\end{tikzpicture}

%% file: FigGammaLDOS.tex

\begin{tikzpicture} 
	\begin{scope}
		\node[](c) at (-0.63, 3.21) {\textbf{(a)}};
		\begin{axis}[
			height=5cm,
			width=5cm,
			xlabel={$\textnormal{Re}(S_{NN})$},
			ylabel={$\textnormal{Im}(S_{NN})$},
			xmin=-1.1,xmax=1.1,
			ymin=-0.6, ymax=1.6,
			xtick={-1,0,1},
			ytick={-1,0,1},
			x label style={at={(axis description cs:0.5,-0.1)}, anchor=south},
			y label style={at={(axis description cs:0.25,0.5)}, anchor=south},
			grid
			]
			\addplot[domain=-1:1, smooth, samples=50] {sqrt(1-x^2)};
			\addplot[black, mark=o] coordinates {(-1,0)};
			\node at (axis cs:-0.6,-0.15) {\footnotesize{$S_{11}=-1$}};
			\addplot[blue, mark=o, only marks] coordinates {(0.9999990286749644, 0.0013937894847786425)};
			\addplot[green, mark=o, only marks] coordinates {(0.9996115452602345, 0.027870388989860127)};
			\addplot[orange, mark=o, only marks] coordinates {(0.9903336905903376, 0.13870537582127598)};
			\addplot[red, mark=o, only marks] coordinates {(-0.9596499840860415, 0.2811972760246104)};
			\addplot[<-,domain=0.1:0.8, smooth, samples=50] {sqrt(0.8-x^2)};
			\node at (axis cs:0.2,0.4) {\footnotesize{Increasing $\Gamma$}};
			\addplot[->,domain=0.1:0.8, smooth, samples=50] {sqrt(1.2-x^2)};
			\node at (axis cs:0.4,1.2) {\footnotesize{Increasing $p$}};
		\end{axis}
	\end{scope}
	\begin{scope}[xshift=3.5cm]
		\begin{axis}[
			height=2.35cm,
			width=5cm,
			yticklabels={,,},
			xlabel={Site Index},
			x label style={at={(axis description cs:0.5,-0.33)}, anchor=south},
			y label style={at={(axis description cs:0.25,0.5)}, anchor=south},
			grid
			]
			\addplot[red] table[x index = {0}, y index = {4}, col sep=comma] {gamma_ldos.csv};
			\node[align=center] at (axis cs:12.5,0.80) {\scriptsize{$\tfrac{\Gamma}{2\pi}{=}125\,\textnormal{GHz}$}};
		\end{axis}
	\end{scope}
	\begin{scope}[xshift=3.5cm, yshift=0.883333cm]
		\begin{axis}[
			height=2.35cm,
			width=5cm,
			xticklabels={,,}, yticklabels={,,},
			x label style={at={(axis description cs:0.5,-0.1)}, anchor=south},
			y label style={at={(axis description cs:0.25,0.5)}, anchor=south},
			grid
			]
			\addplot[orange] table[x index = {0}, y index = {3}, col sep=comma] {gamma_ldos.csv};
			\node[align=center] at (axis cs:34.5,0.80) {\scriptsize{$\tfrac{\Gamma}{2\pi}{=}1.25\,\textnormal{GHz}$}};
		\end{axis}
	\end{scope}
	\begin{scope}[xshift=3.5cm, yshift=1.7666666666cm]
		\begin{axis}[
			height=2.35cm,
			width=5cm,
			xticklabels={,,}, yticklabels={,,},
			x label style={at={(axis description cs:0.5,-0.1)}, anchor=south},
			y label style={at={(axis description cs:0.25,0.5)}, anchor=south},
			grid
			]
			\addplot[green] table[x index = {0}, y index = {2}, col sep=comma] {gamma_ldos.csv};
			\node[align=center] at (axis cs:34.5,0.80) {\scriptsize{$\tfrac{\Gamma}{2\pi}{=}250\,\textnormal{MHz}$}};
		\end{axis}
	\end{scope}
	\begin{scope}[xshift=3.5cm, yshift=2.65cm]
		\begin{axis}[
			height=2.35cm,
			width=5cm,
			xticklabels={,,}, yticklabels={,,},
			x label style={at={(axis description cs:0.5,-0.1)}, anchor=south},
			y label style={at={(axis description cs:0.25,0.5)}, anchor=south},
			grid
			]
			\addplot[blue] table[x index = {0}, y index = {1}, col sep=comma] {gamma_ldos.csv};
			\node[align=center] at (axis cs:34.5,0.80) {\scriptsize{$\tfrac{\Gamma}{2\pi}{=}12.5\,\textnormal{MHz}$}};
		\end{axis}
	\end{scope}
	
	\begin{scope}[yshift=-4cm]
		\node[](c) at (-0.63, 3.1) {\textbf{(b)}};
		\begin{axis}[
			height=4.5cm,
			width=8.5cm,
			xlabel={$\Gamma~(\textnormal{Hz})$},
			ylabel={$\chi$},
			xmode=log,
			ymode=log,
			xmin=12500000,xmax=1250000000000,
			y label style={at={(axis description cs:0.07,0.5)}, anchor=south},
			grid=both
			]
			\addplot[red] table[x index = {0}, y index = {1}, col sep=comma] {gamma_ldos_chis.csv};
		\end{axis}
	\end{scope}
\end{tikzpicture}

%% file: figCqedTranslation.tex
\begin{tikzpicture}[
	circ/.style = {circle, minimum size=#1,
		inner sep=0pt, outer sep=0pt},
	/tikz/circuitikz/bipoles/length=0.5cm
	]
	
	\node[draw, circ="15pt"] (b) at (0,0.9) {\scriptsize{20}};
	\node[draw, circ="11pt"] (a) at (0,0) {\scriptsize{10}};
	\draw[-, line width="1pt"] (b) -- (a);
	\node[draw, circ="20pt"] (b) at (0.9,0) {\scriptsize{11}};
	\draw[-, line width="1pt"] (b) -- (a);
	\node[draw, circ="10pt"] (d) at (1.8,0) {\scriptsize{12}};
	\draw[-, line width="1pt"] (d) -- (b);
	\node[draw, circ="10pt"] (b) at (-0.9,0) {\scriptsize{9}};
	\draw[-, line width="1pt"] (b) -- (a);
	\node[draw, circ="20pt"] (e) at (-1.8,0) {\scriptsize{8}};
	\draw[-, line width="1pt"] (b) -- (e);
	
	\foreach \i in {1,...,4}
	{
		\pgfmathparse{int(2*\i + 11)} 
		\edef\textVal{\pgfmathresult}
		\node[draw, circ="20pt"] (a) at (\i*1.8+0.9,0.0) {\footnotesize \textVal};
		
		\draw[-, line width="3pt"] (a) -- (d) node [pos=0.5, above] {$t_2$};
		
		\pgfmathparse{int(2*\i+12)}
		\edef\textVal{\pgfmathresult}
		\ifnum \i<4
		\node[draw, circ="10pt"] (d) at (\i*1.8+1.8,0.0) {\footnotesize \textVal};
		\draw[-, line width="1pt"] (a) -- (d) node [pos=0.5, above] {$t_1$};
		\fi
	};
	
	\foreach \i in {1,...,4}
	{
		\pgfmathparse{int(9-2*\i)} 
		\edef\textVal{\pgfmathresult}
		\node[draw, circ="10pt"] (a) at (-\i*1.8-0.9,0.0) {\footnotesize \textVal};
		
		\draw[-, line width="3pt"] (a) -- (e) node [pos=0.5, above] {$t_2$};
		
		\pgfmathparse{int(8-2*\i)}
		\edef\textVal{\pgfmathresult}
		\ifnum \i<4
		\node[draw, circ="20pt"] (e) at (-\i*1.8-1.8,0.0) {\footnotesize \textVal};
		\draw[-, line width="1pt"] (a) -- (e) node [pos=0.5, above] {$t_1$};
		\fi
	};

	\node[](c) at (-8.5, 0.9) {\textbf{(a)}};
	
	\begin{scope}[yshift=-1.8cm]
		\node[](c) at (-8.5, 0.9) {\textbf{(b)}};
		\foreach \i in {1,...,19}
		{
			\pgfmathparse{0.9*(\i-1)}
			\edef\posRes{\pgfmathresult}
			\begin{scope}[xshift=\posRes*1cm]
				\coordinate (vLr) at (-7.2,-0.7) {} {} {} {} {} {};
				\pgfmathparse{\i}
				\edef\indM{\pgfmathresult}
				\pgfmathparse{int(\i+1}
				\edef\indN{\pgfmathresult}
				
				\edef\posY{-1.2};
				\newcommand{\opts}{/tikz/circuitikz/bipoles/length=0.4cm}
				\newcommand{\optss}{/tikz/circuitikz/bipoles/length=0.3cm}
				\ifnum \i<10
				\pgfmathparse{int(Mod(\i,2))};
				\ifnum \pgfmathresult=0
				\edef\posY{-0.9};
				\renewcommand{\opts}{/tikz/circuitikz/bipoles/length=0.3cm}
				\else
				\edef\posY{-1.6};
				\renewcommand{\opts}{/tikz/circuitikz/bipoles/length=0.5cm}
				\fi
				
				\ifnum \i<9
				\pgfmathparse{int(Mod(\i+1,2))};
				\ifnum \pgfmathresult=0
				\renewcommand{\optss}{/tikz/circuitikz/bipoles/length=0.53cm}
				\fi
				\fi
				\fi
				\ifnum \i>10
				\pgfmathparse{int(Mod(\i,2))};
				\ifnum \pgfmathresult=1
				\edef\posY{-0.9};
				\renewcommand{\opts}{/tikz/circuitikz/bipoles/length=0.3cm}
				\else
				\edef\posY{-1.6};
				\renewcommand{\opts}{/tikz/circuitikz/bipoles/length=0.5cm}
				\renewcommand{\optss}{/tikz/circuitikz/bipoles/length=0.53cm}
				\fi
				\fi
				
				\ifnum \i < 19
				\draw (-8,-0.7) coordinate (v3) {} to [style/.expanded=\optss, C, l=\footnotesize{$C_{{\indM},{\indN}}$}] (vLr);
				\else
				\coordinate (v3) at (-8.1,-0.7);
				\fi
				
				\edef\posY{-0.9};
				\draw (-8,\posY-0.1) coordinate (v2) {}  to [cute inductor] (-8,\posY-0.5) coordinate (v5) {};
				\draw (-8.25,\posY-0.1) coordinate (v1) {} to [style/.expanded=\opts,capacitor] (-8.25,\posY-0.5) coordinate (v4) {};
				\node (e) at (-8.1,\posY-1.1) {\footnotesize{\indM}};

				\draw  (v1) to[short,-] (-8.25,\posY) -- (-8,\posY) to[short,-] (v2);
				
				\draw (v3) -- (-8.1,-0.7) to[short,-*]  (-8.1,\posY);
				\ifnum \i < 19
				\draw (v3) to[short,-] (-8.1,-0.7) to[short,-*]  (-8.1,\posY);
				\fi
				
				\draw (v4) to[short,-] (-8.25,\posY-0.6) -- (-8,\posY-0.6) to[short,-] (v5);
				\draw[/tikz/circuitikz/bipoles/length=0.8cm] (-8.1,\posY-0.6) to (-8.1,\posY-0.61) node[ground]{};
				
				\ifnum \i=1
					\renewcommand{\opts}{/tikz/circuitikz/bipoles/length=0.3cm}
					\draw ($(-8.1,-0.7) - (0.3,0)$) to [style/.expanded=\opts, C, l=\footnotesize{$c_1$}] (-8.1,-0.7);
					\node[tlinestub,xscale=0.8, yscale=2.0, rotate=180] at ($(-8.1,-0.7) - (0.3,0)$) {};
					\node[] (e) at ($(-8.1,-0.7) + (-0.5,-0.35)$) {\footnotesize{$\Sigma_{1}$}}; 
				\fi
				\ifnum \i=19
					\renewcommand{\opts}{/tikz/circuitikz/bipoles/length=0.3cm}
					\draw ($(-8.1,-0.7) + (0.3,0)$) to [style/.expanded=\opts, C, l_=\footnotesize{}] (-8.1,-0.7);
					\node[tlinestub,xscale=0.8, yscale=2.0] at ($(-8.1,-0.7) + (0.3,0)$) {};
					\node[] (e) at ($(-8.1,-0.7) + (0.2,0.2)$) {\footnotesize{$c_{19}$}}; 
					\node[] (e) at ($(-8.1,-0.7) + (0.5,-0.35)$) {\footnotesize{$\Sigma_{19}$}}; 
				\fi
				
			\end{scope}
		};
		
		\draw (0.2,1) coordinate (v7) {} to [squid, l^=$J_{20}$] (0.2,0.2) coordinate (v6) {};
		\draw (v6) -- (-0.2,0.2) coordinate (v8) {};
		\draw (0,-0.3) coordinate (v10) {} to[/tikz/circuitikz/bipoles/length=0.3cm, capacitor,l=\footnotesize{$C_{10,20}$}] (0,0.2);
		\draw (v7) -- (-0.2,1) coordinate (v9) {};
		\draw (v8) to[capacitor,l=\footnotesize{$C_{20}$}] (v9);
		\draw (v10) -- (0,-0.7);
		\draw[/tikz/circuitikz/bipoles/length=0.8cm] (0,1) -- (0,1.2) to (0.5,1.2) node[ground]{};
	\end{scope}
	
\end{tikzpicture}

%% file: figQuantumDots.tex
\begin{tikzpicture}
	
	\definecolor{babyblueeyes}{rgb}{0.63, 0.79, 0.95}
	
	\foreach \i in {0,...,9}
	{
		\draw  (-\i*0.4-0.05,1.2) rectangle (-\i*0.4+0.05,0.11);
		\draw[fill=babyblueeyes]  (-\i*0.4,0.11+0.08) ellipse (0.05 and 0.07);
		\draw[fill=red]  (-\i*0.4-0.2-0.05,1.2) rectangle (-\i*0.4-0.2+0.05,0.11);
		\draw  (\i*0.4-0.05,1.2) rectangle (\i*0.4+0.05,0.11);
		\draw[fill=babyblueeyes]  (\i*0.4,0.11+0.08) ellipse (0.05 and 0.07);
		\draw[fill=red]  (\i*0.4+0.2-0.05,1.2) rectangle (\i*0.4+0.2+0.05,0.11);
	}
	
	\foreach \i in {1,...,19}
	{
		\node at (-4.0+\i*0.4,0) {\scriptsize\i};
	}
	
	\draw[fill=red]  (-0.2-0.05,-1.2) rectangle (-0.2+0.05,-0.51);
	\draw  (-0.05,-1.2) rectangle (0.05,-0.51);
	\draw[fill=red]  (0.2-0.05,-1.2) rectangle (0.2+0.05,-0.51);
	\draw[fill=babyblueeyes]  (0,-0.51-0.08) ellipse (0.05 and 0.07);
	\node at (0,-0.4) {\scriptsize 20};
	
	\draw (0,-1.45) node[vsourcesquareshape, rotate=-90, /tikz/circuitikz/bipoles/length=0.5cm](srcNode){};
	\draw (srcNode) -- (0,-1.2);

	\begin{scope}
		\draw[fill=red]  (1.4,-0.3) rectangle (1.5,-0.5);
		\node[anchor=west,align=center] at (1.5,-0.4) {Barrier Gate};
	\end{scope}
	\begin{scope}[shift={(0,-0.35)}]
		\draw  (1.4,-0.3) rectangle (1.5,-0.5);
		\node[anchor=west,align=center] at (1.5,-0.42) {Plunger Gate};
	\end{scope}
	\begin{scope}[shift={(0,-0.7)}]
		\draw[fill=babyblueeyes]  (1.45,-0.4) ellipse (0.05 and 0.1);
		\node[anchor=west,align=center] at (1.5,-0.43) {Quantum Dot};
	\end{scope}
	
	\node[align=center] at (-2.1,-0.8) {Gate-Defined\\Quantum Dots};

	\begin{scope}[shift={(0,-2.0)}]
		
		\foreach \i in {1,...,4}
		{
			\draw[fill=yellow]  (-\i*0.8,-2) ellipse (0.05 and 0.05);
			\draw[fill=yellow]  (-\i*0.8-0.32,-2.0+0.1) ellipse (0.05 and 0.05);
			\draw[fill=yellow]  (\i*0.8+0.32,-2) ellipse (0.05 and 0.05);
			\draw[fill=yellow]  (\i*0.8,-2+0.1) ellipse (0.05 and 0.05);
		}
		\draw[fill=yellow]  (0,-2) ellipse (0.05 and 0.05);
		\draw[fill=yellow]  (-0.4,-2+0.1) ellipse (0.05 and 0.05);
		\draw[fill=yellow]  (0.4,-2) ellipse (0.05 and 0.05);
		\foreach \i in {1,...,19}
		{
			\node at (-4.0+\i*0.4,-1.7) {\scriptsize\i};
		}
		\draw[fill=yellow]  (0.0,-2.7) ellipse (0.05 and 0.05);
		
		\draw [fill=yellow] (-0.2,-3.2) rectangle (0.2,-4);
		\draw [fill=yellow]  (-3.5,0.2) rectangle (-1.5,-0.5);
		\draw [fill=yellow]  (1.5,0.2) rectangle (3.5,-0.5);
		\draw [fill=yellow] (-0.25,0.2) rectangle (0.25,-1);
		\node at (0,-2.5) {\scriptsize 20};
		\node[align=center] at (-2.1,-3.2) {P-donor\\Quantum Dots};
				
		\draw[fill=yellow]  (1.45,-2.8) ellipse (0.05 and 0.05);
		\node[anchor=west,align=center] at (1.5,-2.8) {1P Dot};
		\node[anchor=west,align=center] at (1.5,-3.3) {P-donor gate};
		
		\draw [fill=yellow]  (1.4,-3.2) rectangle (1.5,-3.4);
		
		\draw (0,-4.25) node[vsourcesquareshape, rotate=-90, /tikz/circuitikz/bipoles/length=0.5cm](srcNode){};
		\draw (srcNode) -- (0,-4);
	\end{scope}

	\begin{scope}[shift={(0,-7.0)}]
		
		\foreach \i in {1,...,4}
		{
			\draw[fill=yellow]  (-\i*0.8,-2) ellipse (0.05 and 0.05);
			\draw[fill=yellow]  (-\i*0.8-0.32,-2.0+0.05) ellipse (0.05 and 0.05);
			\draw[fill=yellow]  (-\i*0.8-0.32,-2.0-0.05) ellipse (0.05 and 0.05);
			\draw[fill=yellow]  (\i*0.8+0.32,-2) ellipse (0.05 and 0.05);
			\draw[fill=yellow]  (\i*0.8,-2+0.05) ellipse (0.05 and 0.05);
			\draw[fill=yellow]  (\i*0.8,-2-0.05) ellipse (0.05 and 0.05);
		}
		\draw[fill=yellow]  (0,-2) ellipse (0.05 and 0.05);
		\draw[fill=yellow]  (-0.4,-2+0.05) ellipse (0.05 and 0.05);
		\draw[fill=yellow]  (-0.4,-2-0.05) ellipse (0.05 and 0.05);
		\draw[fill=yellow]  (0.4,-2) ellipse (0.05 and 0.05);
		\foreach \i in {1,...,19}
		{
			\node at (-4.0+\i*0.4,-1.7) {\scriptsize\i};
		}
		\draw[fill=yellow]  (0.0,-2.7) ellipse (0.05 and 0.05);
		
		\draw [fill=yellow] (-0.2,-3.2) rectangle (0.2,-4);
		\draw [fill=yellow]  (-3,0.2) rectangle (-1.5,-0.5);
		\draw [fill=yellow]  (1.5,0.2) rectangle (3,-0.5);
		\draw [fill=yellow] (-0.25,0.2) rectangle (0.25,-1);
		\node at (0,-2.5) {\scriptsize 20};
		\node[align=center] at (-2.1,-3.2) {P-donor\\Quantum Dots};
		
		\draw[fill=yellow]  (1.45,-2.8) ellipse (0.05 and 0.05);
		\node[anchor=west,align=center] at (1.5,-2.8) {1P Dot};
		\draw[fill=yellow]  (1.45,-3.3+0.05) ellipse (0.05 and 0.05);
		\draw[fill=yellow]  (1.45,-3.3-0.05) ellipse (0.05 and 0.05);
		\node[anchor=west,align=center] at (1.5,-3.3) {2P Dot};
		\node[anchor=west,align=center] at (1.5,-3.8) {P-donor gate};
		
		\draw [fill=yellow]  (1.4,-3.7) rectangle (1.5,-3.9);
		
		\draw (0,-4.25) node[vsourcesquareshape, rotate=-90, /tikz/circuitikz/bipoles/length=0.5cm](srcNode){};
		\draw (srcNode) -- (0,-4);
	\end{scope}

	\node at (-4.1,1) {\textbf{(a)}};
	\node at (-4.1,-2.2) {\textbf{(b)}};
	\node at (-4.1,-7.2) {\textbf{(c)}};
	\draw (-3.5,-1.65) -- (3.5,-1.65);
	\draw (-3.5,-6.5) -- (3.5,-6.5);
\end{tikzpicture}

%% file: FigSSH.tex
\begin{tikzpicture}[
	circ/.style = {circle, minimum size=#1,
		inner sep=0pt, outer sep=0pt}
	]
	\definecolor{skyblue}{rgb}{0.3686, 0.72549, 1.0}
	
	\begin{scope}
		\node[](c) at (-1, 2.2) {\textbf{(a)}};
		\begin{axis}[
			height=4cm,
			width=8.5cm,
			xlabel={Eigenvalue},
			ylabel={Energy},
			xmin=0, xmax=27,
			ymax=5,
			x label style={at={(axis description cs:0.5,-0.13)}, anchor=south},
			y label style={at={(axis description cs:0.07,0.5)}, anchor=south},
			legend pos=south east,
			grid
			]
			\draw[dashed, fill=gray, fill opacity=0.1] (axis cs: 1.9*1.5+0.425,3.25) rectangle (axis cs: 2.5*1.9+0.95-0.425,4.25);
			\node[align=center] at (axis cs:4.275,2.75) {unit-cell};		\draw[->] (axis cs: 25,1)--(axis cs: 25,2.5);
			\node[align=center] at (axis cs:25,0) {Bloch\\states};
			\draw[->] (axis cs: 9,0)--(axis cs: 13,0);
			\node[align=center] at (axis cs:7,0) {Band\\gap};
			\pgfplotsinvokeforeach{1,...,13}
			{
				\node[draw, circ="5pt"] (a) at (axis cs: #1*1.9,3.75) {};
				\ifnum #1 > 1
				\draw[-, line width=1] (a) -- (d);
				\fi
				\node[draw, circ="5pt"] (d) at (axis cs: #1*1.9+0.95,3.75) {};
				\draw[-, line width=2.5] (a) -- (d);
			}
			\addplot[red, only marks, mark=o] table[x index = {0}, y index = {1}, col sep=comma] {sshOnlyGap.csv};
		\end{axis}
	\end{scope}
	
	\pgfkeys{/tikz/savenumber/.code 2 args={\global\edef#1{#2}}}
	\begin{scope}[yshift=-3.4cm]
		\node[](c) at (-1, 2.2) {\textbf{(b)}};
		\begin{axis}[
			height=4cm,
			width=8.5cm,
			xlabel={Eigenvalue},
			ylabel={Energy},
			xmin=0, xmax=27,
			ymax=5.5,
			x label style={at={(axis description cs:0.5,-0.13)}, anchor=south},
			y label style={at={(axis description cs:0.07,0.5)}, anchor=south},
			legend pos=south east,
			grid
			]
			\node[draw, circ="5pt"] (d) at (axis cs: 0.9,3.75) {};
			\pgfplotsinvokeforeach{1,...,13}
			{
				\node[draw, circ="5pt"] (a) at (axis cs: #1*1.9,3.75) {};
				\draw[-, line width=1] (a) -- (d);
				\node[draw, circ="5pt"] (d) at (axis cs: #1*1.9+0.95,3.75) {};
				\draw[-, line width=2.5] (a) -- (d);
			}
			\addplot[red, only marks, mark=o] table[x index = {0}, y index = {1}, col sep=comma] {sshGapState.csv};
			\addplot[blue, mark=o] coordinates {(13,0)};
			\coordinate (insetSW) at (axis cs:0.95,4.2); 
			\coordinate (insetNE) at (axis cs:25.65,5.25);
			\path
			let
			\p1 = (insetSW),
			\p2 = (insetNE),
			\n1 = {(\x2 - \x1)},
			\n2 = {(\y2 - \y1)}
			in    
			[savenumber={\insetwidth}{\n1},savenumber={\insetheight}{\n2}];
		\end{axis}
		\begin{axis}[
			at={(insetSW)}, anchor=south west,
			name=inset,
			hide axis,
			xmin=0, xmax=26,
			width=\insetwidth, height=\insetheight,
			xticklabels={}, yticklabels={},
			xtick=\empty,ytick=\empty, scale only axis, axis lines = center]
			\addplot[fill=blue, fill opacity=0.5, domain=0:26, samples at={0,...,26}, blue]{exp(-0.25*x) * 0.5*(1-(-1)^(x+1))}\closedcycle;
			\addplot[domain=0:26, smooth, samples=50, dashed, blue] {exp(-0.25*x)};
		\end{axis}  
	\end{scope}
	
	\begin{scope}[yshift=-6.8cm]
		\node[](c) at (-1, 2.2) {\textbf{(c)}};
		\begin{axis}[
			height=4cm,
			width=8.5cm,
			xlabel={Eigenvalue},
			ylabel={Energy},
			xmin=0, xmax=29,
			ymax=5.5,
			x label style={at={(axis description cs:0.5,-0.13)}, anchor=south},
			y label style={at={(axis description cs:0.07,0.5)}, anchor=south},
			legend pos=south east,
			grid
			]
			\pgfplotsinvokeforeach{1,...,13}
			{
				\node[draw, circ="5pt"] (a) at (axis cs: #1*2.0,3.75) {};
				\ifnum #1 > 1
				\draw[-, line width=1] (a) -- (d);
				\fi
				\node[draw, circ="5pt"] (d) at (axis cs: #1*2.0+1.0,3.75) {};
				\draw[-, line width=2.5] (a) -- (d);
				\ifnum #1 = 5
				\node[draw, circ="5pt"] (dd) at (d) {};
				\node[draw, circ="5pt"] (b) at (axis cs: #1*2.0+1.0,4.75) {};
				\fi
				\ifnum #1 = 8
				\node[draw, circ="5pt"] (e) at (a) {};
				\node[draw, circ="5pt"] (f) at (axis cs: #1*2.0,4.75) {};
				\fi
			}
			\addplot[red, only marks, mark=o] table[x index = {0}, y index = {1}, col sep=comma] {sshDoubleSide.csv};
			\addplot[blue, mark=o, only marks] coordinates {(14,0)};
			\addplot[green, mark=o, only marks] coordinates {(15,0)};
			\coordinate (insetSW) at (axis cs:2.0,4.2); 
			\coordinate (insetNE) at (axis cs:27.0,5.25);
			\path
			let
			\p1 = (insetSW),
			\p2 = (insetNE),
			\n1 = {(\x2 - \x1)},
			\n2 = {(\y2 - \y1)}
			in    
			[savenumber={\insetwidth}{\n1},savenumber={\insetheight}{\n2}];
		\end{axis}
		\begin{axis}[
			at={(insetSW)}, anchor=south west,
			name=inset,
			hide axis,
			xmin=0, xmax=25,
			width=\insetwidth, height=\insetheight,
			xticklabels={}, yticklabels={},
			xtick=\empty,ytick=\empty, scale only axis, axis lines = center]
			\addplot[fill=blue, fill opacity=0.5, samples at={0,...,9}, blue]{exp(0.25*(x-9)) * 0.5*(1-(-1)^(x))}\closedcycle;
			\addplot[domain=0:9, smooth, samples=50, dashed, blue] {exp(0.25*(x-9))};
		\end{axis}
		\node[draw, circ="5pt",fill=white] (bb) at (b) {};
		\node[draw, circ="5pt"] (ddd) at (dd) {};
		\draw[-, line width=1] (bb) -- (ddd);
		\begin{axis}[
			at={(insetSW)}, anchor=south west,
			name=inset,
			hide axis,
			xmin=0, xmax=25,
			width=\insetwidth, height=\insetheight,
			xticklabels={}, yticklabels={},
			xtick=\empty,ytick=\empty, scale only axis, axis lines = center]
			\addplot[fill=green, fill opacity=0.5, samples at={14,...,25}, green]{exp(-0.25*(x-14)) * 0.5*(1-(-1)^(x+1))}\closedcycle;
			\addplot[domain=14:25, smooth, samples=50, dashed, green] {exp(-0.25*(x-14))};
		\end{axis}
		\node[draw, circ="5pt", fill=white] (ee) at (e) {};
		\node[draw, circ="5pt", fill=white] (ff) at (f) {};
		\draw[-, line width=1] (ee) -- (ff);
	\end{scope}
\end{tikzpicture}

%% file: fig_straight.tex
\newcommand{\combFour}[4]{

	\node[draw, circ=8pt] (b) at (0, 0) {};
	\node[below] at (b.south) {$V_\textnormal{C}$};
	\foreach \i in {1,...,3}
	{
		\pgfmathparse{#1 ? "10pt" : "6pt"}
		\edef\Usize{\pgfmathresult}
		\node[draw, circ=\Usize] (a) at (-1.2*\i+0.6, 0) {};
		\node[below] at (a.south) {\ifnum #1=1 $V_\textnormal{R}$ \else $V_\textnormal{L}$ \fi};

		\ifnum\i=1
		\draw[-, line width=2pt] (b) -- (a) node [pos=0.5, above] {\ifnum\i=1 \ifnum #3=0 $t_\textnormal{L}$ \else $t_\textnormal{R}$ \fi \fi};
		\else
		\draw[-, line width=3pt] (b) -- (a) node [pos=0.5, above] {\ifnum\i=2 $t_2$ \fi};
		\fi
		
		\pgfmathparse{#1 ? "6pt" : "10pt"}
		\edef\Usize{\pgfmathresult}
		\node[draw, circ=\Usize] (b) at (-1.2*\i, 0) {};
		\node[below] at (b.south) {\ifnum #1=1 $V_\textnormal{L}$ \else $V_\textnormal{R}$ \fi};
		\pgfmathparse{#3 ? "1pt" : "3pt"}
		\edef\tsize{\pgfmathresult}
		\draw[-, line width=1pt] (b) -- (a);
	}
	\pgfmathparse{#3 ? "3pt" : "1pt"}
	\edef\tsize{\pgfmathresult}
	\draw[-, line width=\tsize] (b) -- ($ (b) - (0.3,0) $);
	\node[draw, circ=8pt] (b) at (0, 0) {};
	\foreach \i in {1,...,3}
	{
		\pgfmathparse{#2 ? "10pt" : "6pt"}
		\edef\Usize{\pgfmathresult}
		\node[draw, circ=\Usize] (a) at (1.2*\i-0.6, 0) {};
		\node[below] at (a.south) {\ifnum #2=1 $V_\textnormal{R}$ \else $V_\textnormal{L}$ \fi};
		
		\ifnum\i=1
		\draw[-, line width=2pt] (b) -- (a) node [pos=0.5, above] {\ifnum\i=1 \ifnum #3=0 $t_\textnormal{R}$ \else $t_\textnormal{R}$ \fi \fi};
		\else
		\draw[-, line width=3pt] (b) -- (a) node [pos=0.5, above] {};
		\fi
		
		\pgfmathparse{#2 ? "6pt" : "10pt"}
		\edef\Usize{\pgfmathresult}
		\node[draw, circ=\Usize] (b) at (1.2*\i, 0) {};
		\node[below] at (b.south) {\ifnum #2=1 $V_\textnormal{L}$ \else $V_\textnormal{R}$ \fi};
		\pgfmathparse{#3 ? "1pt" : "3pt"}
		\edef\tsize{\pgfmathresult}
		\draw[-, line width=1pt] (b) -- (a) node [pos=0.5, above] {\ifnum\i=2 $t_1$ \fi};
	}
	\pgfmathparse{#4 ? "3pt" : "1pt"}
	\edef\tsize{\pgfmathresult}
	\draw[-, line width=\tsize] (b) -- ($ (b) + (0.3,0) $);
	\draw[draw=black, dashed] (-0.9,0.6) rectangle (0.9,-0.8);
}

\begin{tikzpicture}[
	circ/.style = {circle, minimum size=#1,
		inner sep=0pt, outer sep=0pt}
	]
	
	\combFour{0}{1}{0}{1};
	
\end{tikzpicture}

%% file: fig_UdomTc.tex
\newcommand{\combFour}[4]{

\node[draw, circ=8pt] (b) at (0, 0) {};
\node[below] at (b.south) {$V_\textnormal{C}$};
\node[draw, circ=8pt] (q) at (0, 0.9) {};
\node[right] at (q.east) {$V_\textnormal{Q}$};
\draw[-, line width=0.5pt] (b) -- (q) node [pos=0.5, right]{$t_\textnormal{Q}$};
\foreach \i in {1,...,3}
{
	\pgfmathparse{#1 ? "10pt" : "6pt"}
	\edef\Usize{\pgfmathresult}
	\node[draw, circ=\Usize] (a) at (-1.2*\i+0.6, 0) {};
	\node[below] at (a.south) {\ifnum #1=1 $V_\textnormal{R}$ \else $V_\textnormal{L}$ \fi};
	\pgfmathparse{#3 ? "3pt" : "1pt"}
	\edef\tsize{\pgfmathresult}
	\draw[-, line width=\tsize] (b) -- (a) node [pos=0.5, above] {\ifnum\i=2 \ifnum #3=0 $t_\textnormal{L}$ \else $t_\textnormal{R}$ \fi \fi};
	\pgfmathparse{#1 ? "6pt" : "10pt"}
	\edef\Usize{\pgfmathresult}
	\node[draw, circ=\Usize] (b) at (-1.2*\i, 0) {};
	\node[below] at (b.south) {\ifnum #1=1 $V_\textnormal{L}$ \else $V_\textnormal{R}$ \fi};
	\pgfmathparse{#3 ? "1pt" : "3pt"}
	\edef\tsize{\pgfmathresult}
	\draw[-, line width=\tsize] (b) -- (a);
}
\pgfmathparse{#3 ? "3pt" : "1pt"}
\edef\tsize{\pgfmathresult}
\draw[-, line width=\tsize] (b) -- ($ (b) - (0.3,0) $);
\node[draw, circ=8pt] (b) at (0, 0) {};
\foreach \i in {1,...,3}
{
	\pgfmathparse{#2 ? "10pt" : "6pt"}
	\edef\Usize{\pgfmathresult}
	\node[draw, circ=\Usize] (a) at (1.2*\i-0.6, 0) {};
	\node[below] at (a.south) {\ifnum #2=1 $V_\textnormal{R}$ \else $V_\textnormal{L}$ \fi};
	\pgfmathparse{#4 ? "3pt" : "1pt"}
	\edef\tsize{\pgfmathresult}
	\draw[-, line width=\tsize] (b) -- (a) node [pos=0.5, above] {\ifnum\i=2 \ifnum #4=0 $t_\textnormal{L}$ \else $t_\textnormal{R}$ \fi \fi};
	\pgfmathparse{#2 ? "6pt" : "10pt"}
	\edef\Usize{\pgfmathresult}
	\node[draw, circ=\Usize] (b) at (1.2*\i, 0) {};
	\node[below] at (b.south) {\ifnum #2=1 $V_\textnormal{L}$ \else $V_\textnormal{R}$ \fi};
	\pgfmathparse{#4 ? "1pt" : "3pt"}
	\edef\tsize{\pgfmathresult}
	\draw[-, line width=\tsize] (b) -- (a);
}
\pgfmathparse{#4 ? "3pt" : "1pt"}
\edef\tsize{\pgfmathresult}
\draw[-, line width=\tsize] (b) -- ($ (b) + (0.3,0) $);
\draw[draw=black, dashed] (-0.9,1.25) rectangle (0.9,-0.8);
}

\begin{tikzpicture}[
circ/.style = {circle, minimum size=#1,
              inner sep=0pt, outer sep=0pt}
]

\combFour{0}{1}{0}{1};

\end{tikzpicture}

%% file: fig_plotRM_UdomTc.tex
\newcommand{\miniFour}[4]{
	\node[draw, circ=5pt, fill=#4] (b) at (1.2,0.7) {};
	\node[draw, circ=5pt, fill=#3] (q) at (1.2,1.1) {};
	\draw[-, line width=0.5pt] (b) -- (q) node [pos=0.5, right]{};
	\node[draw, circ=7pt, fill=#2] (a) at (1.6,0.7) {};
	\coordinate[] (c) at (1.9,0.7) {} {};
	\draw[-, line width=0.5pt] (c) -- (a) node [pos=0.5, right]{};
	\draw[-, line width=0.5pt] (b) -- (a) node [pos=0.5, right]{};
	\node[draw, circ=3pt, fill=#1] (a) at (0.8,0.7) {};
	\coordinate[] (c) at (0.6,0.7) {} {};
	\draw[-, line width=0.5pt] (c) -- (a) node [pos=0.5, right]{};
	\draw[-, line width=0.5pt] (b) -- (a) node [pos=0.5, right]{};
}

\begin{tikzpicture}[
	circ/.style = {circle, minimum size=#1,
		inner sep=0pt, outer sep=0pt}
	]
	
	\begin{axis}[
		height=5cm,
		width=9cm,
		xmin=-0.2, xmax = 0.2,
		xlabel=$V_\textnormal{Q}$,
		ylabel=$E$,
		x label style={at={(axis description cs:0.5,-0.1)},anchor=south},
		y label style={at={(axis description cs:0.1,0.5)}, anchor=south},
		yticklabels={,,},
		xtick={-0.1,0,0.1},
		xticklabels={$V_\textnormal{L}$,,$V_\textnormal{R}$},
		ytick={-0.1,0,0.1},
		yticklabels={$V_\textnormal{L}$,,$V_\textnormal{R}$},
		grid
		]
		\addplot [black] table[mark=none, x index = {0},y index = {5}]{plotRM_UdomTc.tsv} node [pos=0.25](vArrw0) {};
		\addplot [black] table[mark=none, x index = {0},y index = {6}]{plotRM_UdomTc.tsv};
		\addplot [black] table[mark=none, x index = {0},y index = {7}]{plotRM_UdomTc.tsv};
	\end{axis}
	\node [font=\scriptsize] at (0.7,0.3) {$\left|{\textnormal{Q}}\right\rangle$};
	\node [font=\scriptsize] at (0.7,1.2) {$\left|{\textnormal{L}}\right\rangle$};
	\node [font=\scriptsize] at (3,0.8) {$\left|{\textnormal{L}}\right\rangle$};
	\node [font=\scriptsize] at (3.6,1.9) {$\left|{\textnormal{Q}}\right\rangle$};
	\node [font=\scriptsize] at (4.8,2.6) {$\left|{\textnormal{R}}\right\rangle$};
	\node [font=\scriptsize] at (6.6,2.22) {$\left|{\textnormal{R}}\right\rangle$};
	\node [font=\scriptsize] at (6.6,3.05) {$\left|{\textnormal{Q}}\right\rangle$};
	
	\draw [->] (1.85,0.7) -- (1.85,0.9015);
	\draw [->] (1.85,1.3735) -- (1.85,1.0523);
	\node at (1.87,0.4) {$\frac{t_\textnormal{Q}^2+t_\textnormal{L}^2}{V_\textnormal{C}-V_\textnormal{L}}$};
	
	\draw [->] (5.55,2.0864) -- (5.55,2.35);
	\draw [->] (5.55,2.75) -- (5.55,2.5);
	\node at (5.53,3.05) {$\frac{t_\textnormal{Q}^2+t_\textnormal{R}^2}{V_\textnormal{C}-V_\textnormal{R}}$};

\begin{scope}[xshift=-0.2cm, yshift=0.75cm]
	\begin{axis}[at={(0.2cm,0.7cm)}, height=0.5cm, width=2cm, anchor=south west, scale only axis, hide axis, xmin=-4,xmax=4,ymin=0,ymax=1, xtick=\empty, ytick=\empty] 
		\addplot[shade, left color=white, right color=blue, draw=none, samples=100, domain=-4:0] {exp(-abs(0.5*x))} \closedcycle;
	\end{axis}
	\miniFour{white}{black}{white}{black};
\end{scope}
\begin{scope}[xshift=2.5cm, yshift=0.45cm]
	\begin{axis}[at={(0.2cm,0.7cm)}, height=0.5cm, width=2cm, anchor=south west, scale only axis, hide axis, xmin=-4,xmax=4,ymin=0,ymax=1, xtick=\empty, ytick=\empty] 
		\addplot[shade, left color=white, right color=green, draw=none, samples=100, domain=-4:0] {exp(-abs(4*x))} \closedcycle;
	\end{axis}
	\begin{axis}[at={(0.2cm,0.7cm)}, height=0.5cm, width=2cm, anchor=south west, scale only axis, hide axis, xmin=-4,xmax=4,ymin=0,ymax=1, xtick=\empty, ytick=\empty] 
		\addplot[shade, left color=green, right color=white, draw=none, samples=100, domain=0:4] {exp(-abs(4*x))} \closedcycle;
	\end{axis}
	\miniFour{black}{black}{white}{gray};
\end{scope}
\begin{scope}[xshift=5.1cm, yshift=0.9cm]
	\begin{axis}[at={(0.2cm,0.7cm)}, height=0.5cm, width=2cm, anchor=south west, scale only axis, hide axis, xmin=-4,xmax=4,ymin=0,ymax=1, xtick=\empty, ytick=\empty] 
	\addplot[shade, left color=red, right color=white, draw=none, samples=100, domain=0:4] {exp(-abs(0.5*x))} \closedcycle;
	\end{axis}
	\miniFour{black}{white}{white}{black};
\end{scope}
\end{tikzpicture}

%% file: fig_comb4.tex
\newcommand{\combFour}[4]{

\node[draw, circ=8pt] (b) at (0, 0) {};
\node[below] at (b.south) {$V_C$};
\node[draw, circ=8pt] (q) at (0, 0.9) {};
\node[right] at (q.east) {$V_Q$};
\draw[-, line width=0.5pt] (b) -- (q) node [pos=0.5, right]{$t_\textnormal{Q}$};
\foreach \i in {1,...,3}
{
	\pgfmathparse{#1 ? "10pt" : "6pt"}
	\edef\Usize{\pgfmathresult}
	\node[draw, circ=\Usize] (a) at (-1.2*\i+0.6, 0) {};
	\node[below] at (a.south) {\ifnum #1=1 $V_2$ \else $V_1$ \fi};
	\pgfmathparse{#3 ? "3pt" : "1pt"}
	\edef\tsize{\pgfmathresult}
	\draw[-, line width=\tsize] (b) -- (a) node [pos=0.5, above] {\ifnum\i=2 \ifnum #3=0 $t_1$ \else $t_2$ \fi \fi};
	\pgfmathparse{#1 ? "6pt" : "10pt"}
	\edef\Usize{\pgfmathresult}
	\node[draw, circ=\Usize] (b) at (-1.2*\i, 0) {};
	\node[below] at (b.south) {\ifnum #1=1 $V_1$ \else $V_2$ \fi};
	\pgfmathparse{#3 ? "1pt" : "3pt"}
	\edef\tsize{\pgfmathresult}
	\draw[-, line width=\tsize] (b) -- (a);
}
\pgfmathparse{#3 ? "3pt" : "1pt"}
\edef\tsize{\pgfmathresult}
\draw[-, line width=\tsize] (b) -- ($ (b) - (0.3,0) $);
\node[draw, circ=8pt] (b) at (0, 0) {};
\foreach \i in {1,...,3}
{
	\pgfmathparse{#2 ? "10pt" : "6pt"}
	\edef\Usize{\pgfmathresult}
	\node[draw, circ=\Usize] (a) at (1.2*\i-0.6, 0) {};
	\node[below] at (a.south) {\ifnum #2=1 $V_2$ \else $V_1$ \fi};
	\pgfmathparse{#4 ? "3pt" : "1pt"}
	\edef\tsize{\pgfmathresult}
	\draw[-, line width=\tsize] (b) -- (a) node [pos=0.5, above] {\ifnum\i=2 \ifnum #4=0 $t_1$ \else $t_2$ \fi \fi};
	\pgfmathparse{#2 ? "6pt" : "10pt"}
	\edef\Usize{\pgfmathresult}
	\node[draw, circ=\Usize] (b) at (1.2*\i, 0) {};
	\node[below] at (b.south) {\ifnum #2=1 $V_1$ \else $V_2$ \fi};
	\pgfmathparse{#4 ? "1pt" : "3pt"}
	\edef\tsize{\pgfmathresult}
	\draw[-, line width=\tsize] (b) -- (a);
}
\pgfmathparse{#4 ? "3pt" : "1pt"}
\edef\tsize{\pgfmathresult}
\draw[-, line width=\tsize] (b) -- ($ (b) + (0.3,0) $);
\draw[draw=black, dashed] (-0.9,1.25) rectangle (0.9,-0.8);
}

\begin{tikzpicture}[
circ/.style = {circle, minimum size=#1,
              inner sep=0pt, outer sep=0pt}
]

\begin{scope}[shift={(0,0)}]
\combFour{0}{1}{0}{1};
\end{scope}
\begin{scope}[shift={(0,-2.5)}]
\combFour{1}{0}{0}{1};
\end{scope}
\begin{scope}[shift={(0,-5)}]
\combFour{0}{0}{0}{1};
\end{scope}
\begin{scope}[shift={(0,-7.5)}]
\combFour{0}{1}{0}{0};
\end{scope}

\end{tikzpicture}

%% file: FigNumPeaks.tex

	\begin{tikzpicture} 
		\begin{scope}
			\begin{axis}[
				height=4cm,
				width=5cm,
				xlabel={Eigenvalue},
				ylabel=$E\,\textnormal{(GHz)}$,
				xtick={0,10,20},
				x label style={at={(axis description cs:0.5,-0.1)}, anchor=south},
				y label style={at={(axis description cs:0.2,0.5)}, anchor=south},
				grid
				]
				\addplot[red, only marks, mark=x] table[x index = {0}, y index = {1}, col sep=comma] {peaks25.csv};
				\addplot[blue, only marks, mark=x] table[x index = {0}, y index = {2}, col sep=comma] {peaks25.csv};
				\node[align=center] at (axis cs:6,7.22) {$C_\textnormal{int}=22\,\textnormal{fF}$\\$C_\textnormal{res}=467\,\textnormal{fF}$};
			\end{axis}
		\end{scope}
		\begin{scope}[xshift=3.5cm]
			\begin{axis}[
				height=4cm,
				width=5cm,
				xlabel={Eigenvalue},
				ylabel near ticks, yticklabel pos=right,
				x label style={at={(axis description cs:0.5,-0.1)}, anchor=south},
				xtick={0,10,20},
				grid
				]
				\addplot[red, only marks, mark=x] table[x index = {0}, y index = {1}, col sep=comma] {peaks10.csv};
				\addplot[blue, only marks, mark=x] table[x index = {0}, y index = {2}, col sep=comma] {peaks10.csv};
				\node[align=center] at (axis cs:6,7.088) {$C_\textnormal{int}=9\,\textnormal{fF}$\\$C_\textnormal{res}=497\,\textnormal{fF}$};
			\end{axis}
		\end{scope}
	\end{tikzpicture}

%% file: FigNumZgreen.tex

\begin{tikzpicture} 
	\begin{scope}
		\begin{axis}[
			height=3cm,
			width=8.5cm,
			ylabel=$\textnormal{S}_{19,1}\,\textnormal{(dB)}$,
			xmin=6.7, xmax=7.3,
			ymin=-80, ymax=0,
			y label style={at={(axis description cs:0.07,0.5)}, anchor=south},
			grid
			]
			\addplot[blue] table[x index = {0}, y index = {1}, col sep=comma] {zGreen25ideal.csv};
			\addplot[red] table[x index = {0}, y index = {1}, col sep=comma] {zGreen25numer.csv};
			\node[align=center] at (axis cs:6.85,-60) {$C_\textnormal{int}=22\,\textnormal{fF}$\\$C_\textnormal{res}=467\,\textnormal{fF}$};
		\end{axis}
	\end{scope}
	\begin{scope}[yshift=-2.1cm]
		\begin{axis}[
			height=3cm,
			width=8.5cm,
			xlabel={Frequency\,(GHz)},
			ylabel=$\textnormal{S}_{19,1}\,\textnormal{(dB)}$,
			xmin=6.87, xmax=7.12,
			ymin=-80, ymax=0,
			x label style={at={(axis description cs:0.5,-0.2)}, anchor=south},
			y label style={at={(axis description cs:0.07,0.5)}, anchor=south},
			grid
			]
			\addplot[blue] table[x index = {0}, y index = {1}, col sep=comma] {zGreen10ideal.csv};
			\addplot[red] table[x index = {0}, y index = {1}, col sep=comma] {zGreen10numer.csv};
			\node[align=center] at (axis cs:6.935,-60) {$C_\textnormal{int}=9\,\textnormal{fF}$\\$C_\textnormal{res}=497\,\textnormal{fF}$};
		\end{axis}
	\end{scope}
\end{tikzpicture}

%% file: figAppenGreen.tex
\begin{tikzpicture}[
	circ/.style = {circle, minimum size=#1,
		inner sep=0pt, outer sep=0pt, fill=white}
	]
	
	\node[tlinestub,rotate=180,xscale=0.4] at (-3.7,0) {};
	\node[tlinestub,xscale=0.4] at (3.8,0) {};
	
	\node[draw, circ="12pt"] (b) at (0,1) {};
	\node[draw, circ="10pt"] (a) at (0,0) {\scriptsize{$M$}};
	\draw[-, line width="1pt"] (b) -- (a);
	\node[draw, circ="12pt"] (d) at (0.6,0) {};
	\draw[-, line width="1pt"] (d) -- (a);
	\node[draw, circ="8pt"] (b) at (-0.6,0) {};
	\draw[-, line width="1pt"] (b) -- (a);
	
	\node[draw, circ="12pt"] (c) at (-1.2,0) {};
	\draw[-, line width="1pt"] (b) -- (c);
	\node[draw, circ="8pt"] (b) at (-1.8,0) {};
	\draw[-, line width="3pt"] (b) -- (c);
	\draw[-, line width="1pt"] (b) -- (-2.2,0);
	\node[draw, circ="12pt"] (c) at (-3,0) {};
	\node[draw, circ="8pt"] (b) at (-3.6,0) {\scriptsize{$1$}};
	\draw[-, line width="3pt"] (b) -- (c);
	\draw[-, line width="1pt"] (c) -- (-2.6,0);

	\node[draw, circ="8pt"] (c) at (1.2,0) {};
	\draw[-, line width="1pt"] (d) -- (c);
	\node[draw, circ="12pt"] (b) at (1.8,0) {};
	\draw[-, line width="3pt"] (b) -- (c);
	\draw[-, line width="1pt"] (b) -- (2.2,0);
	\node[draw, circ="8pt"] (c) at (3,0) {};
	\node[draw, circ="12pt"] (b) at (3.6,0) {\scriptsize{$N$}};
	\draw[-, line width="3pt"] (b) -- (c);
	\draw[-, line width="1pt"] (c) -- (2.6,0);

\node at (-4,-0.4) {\footnotesize{$\Gamma_1$}};
\node at (4.15,-0.4) {\footnotesize{$\Gamma_N$}};
	
\node at (0.7,-0.4) {\footnotesize{$V$}};
\node at (-0.5,-0.4) {\footnotesize{$-V$}};
\node at (0,-0.4) {\footnotesize{$V_M$}};
\node at (-0.3,0.25) {$t_L$};
\node at (0.27,0.25) {$t_R$};
\node at (-0.2,0.6) {$t_Q$};
\node at (0,1.4) {$V_Q$};

\node at (-3.5,-0.4) {\footnotesize{$-V$}};
\node at (-3,-0.4) {\footnotesize{$V$}};
\node at (-1.8,-0.4) {\footnotesize{$-V$}};
\node at (-1.2,-0.4) {\footnotesize{$V$}};
\node at (-0.85,0.25) {$t_1$};
\node at (-1.55,0.25) {$t_2$};
\node at (-2.05,0.25) {$t_1$};
\node at (-2.65,0.25) {$t_1$};
\node at (-3.35,0.25) {$t_2$};

\draw (-3.7,0.3) -- (-3.7,0.5) -- (-1.1,0.5) -- (-1.1,0.3);
\node at (-2.4,0.7) {$p$ pairs};
\node at (-2.4,0) {...};

\node at (3.6,-0.4) {\footnotesize{$V$}};
\node at (3,-0.4) {\footnotesize{$-V$}};
\node at (1.8,-0.4) {\footnotesize{$V$}};
\node at (1.2,-0.4) {\footnotesize{$-V$}};
\node at (0.92,0.25) {$t_1$};
\node at (1.45,0.25) {$t_2$};
\node at (2.15,0.25) {$t_1$};
\node at (2.75,0.25) {$t_1$};
\node at (3.25,0.25) {$t_2$};

\draw (3.7,0.3) -- (3.7,0.5) -- (1.1,0.5) -- (1.1,0.3);
\node at (2.4,0.7) {$p$ pairs};
\node at (2.4,0) {...};

\end{tikzpicture}